\documentclass[onecolumn,aps,superscriptaddress]{revtex4-2}
\usepackage{times}
\usepackage{geometry}
\usepackage[utf8]{inputenc} 
\usepackage{enumitem,amssymb}
\usepackage{ragged2e}
\usepackage{graphicx}
\usepackage{comment}
\usepackage[usenames]{xcolor} 
\definecolor{xlinkcolor}{cmyk}{1,1,0,0}
\usepackage{url}
\usepackage{hyperref}
  \pdfoutput=1 
\hypersetup{pdfauthor={Name},
 colorlinks=true,    
 linkcolor=xlinkcolor,     
 citecolor=xlinkcolor,     
 filecolor=xlinkcolor,  
 urlcolor=xlinkcolor,      
 final=true
}
\PassOptionsToPackage{unicode}{hyperref}
\PassOptionsToPackage{naturalnames}{hyperref}
\usepackage{lineno}  
\usepackage{amsmath}
\usepackage{amssymb}
\usepackage{eurosym}
\usepackage{epsfig}
\usepackage{ifpdf}
\usepackage{subfig}
\def\beq{\begin{equation}}
\def\eeq#1{\label{#1}\end{equation}}
\def\eeqn{\end{equation}}

\newenvironment{Eqnarray}%
   {\arraycolsep 0.14em\begin{eqnarray}}{\end{eqnarray}}
\def\beqa{\begin{Eqnarray}}
\def\eeqa#1{\label{#1}\end{Eqnarray}}
\def\eeqan{\end{Eqnarray}}

\let\bar=\overbar

\def\lsim{\mathrel{\raise.3ex\hbox{$<$\kern-.75em\lower1ex\hbox{$\sim$}}}}
\def\gsim{\mathrel{\raise.3ex\hbox{$>$\kern-.75em\lower1ex\hbox{$\sim$}}}}

\def\del{\partial}
\def\Dslash{\not{\hbox{\kern-4pt $D$}}}
\def\dslash{\not{\hbox{\kern-2pt $\del$}}}
\def\pslash{\not{\hbox{\kern-2pt $p$}}}
\def\ETmiss{\not{\hbox{\kern-4pt $E$}}_T}

\def\Dlr{\mathrel{\raise1.5ex\hbox{$\leftrightarrow$\kern-1em\lower1.5ex\hbox{$D$}}}}

\def\MSB{{\bar{M \kern -2pt S}}}
\def\msb{{\bar{\scriptsize M \kern -1pt S}}}

\def\drb{{\bar{\scriptsize D \kern -1pt R}}}

\newcommand\snowmass{\begin{center}\rule[-0.2in]{\hsize}{0.01in}\\\rule{\hsize}{0.01in}\\
\vskip 0.1in Submitted to the  Proceedings of the US Community Study\\ 
on the Future of Particle Physics (Snowmass 2021)\\ 
\rule{\hsize}{0.01in}\\\rule[+0.2in]{\hsize}{0.01in} \end{center}}

\begin{document}


\title{\includegraphics[width=0.4\textwidth]{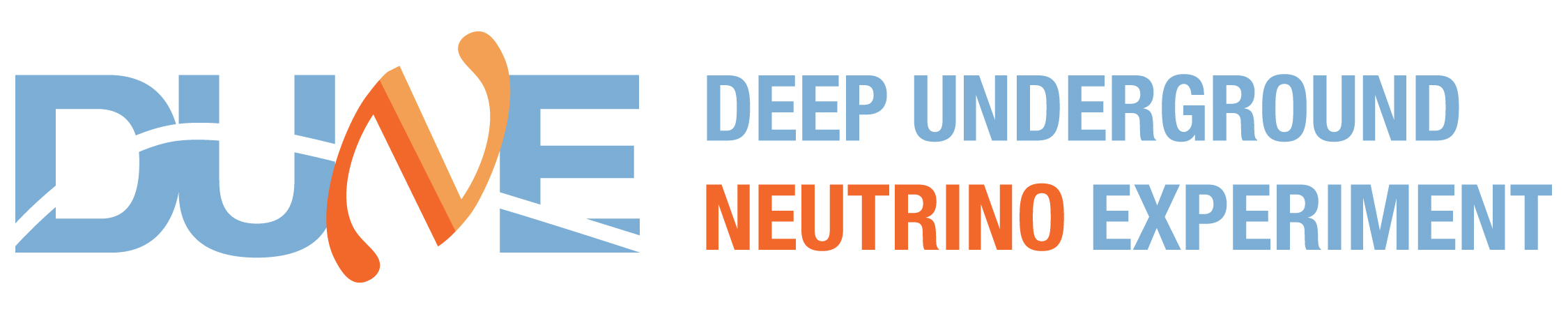} \\ \vspace{1in} A Gaseous Argon-Based Near Detector to Enhance the Physics Capabilities of DUNE \\ \vspace{0.5in} Authors}







\newcommand{\Abilene}{Abilene Christian University, Abilene, TX 79601, USA}
\newcommand{\Albanysuny}{University of Albany, SUNY, Albany, NY 12222, USA}
\newcommand{\Amsterdam}{University of Amsterdam, NL-1098 XG Amsterdam, The Netherlands}
\newcommand{\Antalya}{Antalya Bilim University, 07190 D{\"o}{\c{s}}emealt{\i}/Antalya, Turkey}
\newcommand{\Antananarivo}{University of Antananarivo, Antananarivo 101, Madagascar}
\newcommand{\AntonioNarino}{Universidad Antonio Nari{\~n}o, Bogot{\'a}, Colombia}
\newcommand{\Argonne}{Argonne National Laboratory, Argonne, IL 60439, USA}
\newcommand{\Arizona}{University of Arizona, Tucson, AZ 85721, USA}
\newcommand{\Asuncion}{Universidad Nacional de Asunci{\'o}n, San Lorenzo, Paraguay}
\newcommand{\Athens}{University of Athens, Zografou GR 157 84, Greece}
\newcommand{\Atlantico}{Universidad del Atl{\'a}ntico, Barranquilla, Atl{\'a}ntico, Colombia}
\newcommand{\Augustana}{Augustana University, Sioux Falls, SD 57197, USA}
\newcommand{\Basel}{University of Basel, CH-4056 Basel, Switzerland}
\newcommand{\Bern}{University of Bern, CH-3012 Bern, Switzerland}
\newcommand{\Beykent}{Beykent University, Istanbul, Turkey}
\newcommand{\Birmingham}{University of Birmingham, Birmingham B15 2TT, United Kingdom}
\newcommand{\BolognaUniversity}{Universit{\`a} del Bologna, 40127 Bologna, Italy}
\newcommand{\Boston}{Boston University, Boston, MA 02215, USA}
\newcommand{\Bristol}{University of Bristol, Bristol BS8 1TL, United Kingdom}
\newcommand{\Brookhaven}{Brookhaven National Laboratory, Upton, NY 11973, USA}
\newcommand{\Bucharest}{University of Bucharest, Bucharest, Romania}
\newcommand{\CBPF}{Centro Brasileiro de Pesquisas F\'isicas, Rio de Janeiro, RJ 22290-180, Brazil}
\newcommand{\CEASaclay}{IRFU, CEA, Universit{\'e} Paris-Saclay, F-91191 Gif-sur-Yvette, France}
\newcommand{\CERN}{CERN, The European Organization for Nuclear Research, 1211 Meyrin, Switzerland}
\newcommand{\CIEMAT}{CIEMAT, Centro de Investigaciones Energ{\'e}ticas, Medioambientales y Tecnol{\'o}gicas, E-28040 Madrid, Spain}
\newcommand{\CUSB}{Central University of South Bihar, Gaya, 824236, India }
\newcommand{\CalBerkeley}{University of California Berkeley, Berkeley, CA 94720, USA}
\newcommand{\CalDavis}{University of California Davis, Davis, CA 95616, USA}
\newcommand{\CalIrvine}{University of California Irvine, Irvine, CA 92697, USA}
\newcommand{\CalLosangeles}{University of California Los Angeles, Los Angeles, CA 90095, USA}
\newcommand{\CalRiverside}{University of California Riverside, Riverside CA 92521, USA}
\newcommand{\CalSantabarbara}{University of California Santa Barbara, Santa Barbara, California 93106 USA}
\newcommand{\Caltech}{California Institute of Technology, Pasadena, CA 91125, USA}
\newcommand{\Cambridge}{University of Cambridge, Cambridge CB3 0HE, United Kingdom}
\newcommand{\Campinas}{Universidade Estadual de Campinas, Campinas - SP, 13083-970, Brazil}
\newcommand{\CataniaUniversitadi}{Universit{\`a} di Catania, 2 - 95131 Catania, Italy}
\newcommand{\Catolica}{Universidad Cat{\'o}lica del Norte, Antofagasta, Chile}
\newcommand{\Charles}{Institute of Particle and Nuclear Physics of the Faculty of Mathematics and Physics of the Charles University, 180 00 Prague 8, Czech Republic }
\newcommand{\Chicago}{University of Chicago, Chicago, IL 60637, USA}
\newcommand{\ChungAng}{Chung-Ang University, Seoul 06974, South Korea}
\newcommand{\Cincinnati}{University of Cincinnati, Cincinnati, OH 45221, USA}
\newcommand{\Cinvestav}{Centro de Investigaci{\'o}n y de Estudios Avanzados del Instituto Polit{\'e}cnico Nacional (Cinvestav), Mexico City, Mexico}
\newcommand{\Colima}{Universidad de Colima, Colima, Mexico}
\newcommand{\ColoradoBoulder}{University of Colorado Boulder, Boulder, CO 80309, USA}
\newcommand{\ColoradoState}{Colorado State University, Fort Collins, CO 80523, USA}
\newcommand{\Columbia}{Columbia University, New York, NY 10027, USA}
\newcommand{\Cti}{Centro de Tecnologia da Informacao Renato Archer, Amarais - Campinas, SP - CEP 13069-901}
\newcommand{\CzechAcademyofSciences}{Institute of Physics, Czech Academy of Sciences, 182 00 Prague 8, Czech Republic}
\newcommand{\CzechTechnical}{Czech Technical University, 115 19 Prague 1, Czech Republic}
\newcommand{\DakotaState}{Dakota State University, Madison, SD 57042, USA}
\newcommand{\Dallas}{University of Dallas, Irving, TX 75062-4736, USA}
\newcommand{\DannecyleVieux}{Laboratoire d{\textquoteright}Annecy de Physique des Particules, Univ. Grenoble Alpes, Univ. Savoie Mont Blanc, CNRS, LAPP-IN2P3, 74000 Annecy, France}
\newcommand{\Daresbury}{Daresbury Laboratory, Cheshire WA4 4AD, United Kingdom}
\newcommand{\Drexel}{Drexel University, Philadelphia, PA 19104, USA}
\newcommand{\Duke}{Duke University, Durham, NC 27708, USA}
\newcommand{\Durham}{Durham University, Durham DH1 3LE, United Kingdom}
\newcommand{\EIA}{Universidad EIA, Envigado, Antioquia, Colombia}
\newcommand{\ETH}{ETH Zurich, Zurich, Switzerland}
\newcommand{\Edinburgh}{University of Edinburgh, Edinburgh EH8 9YL, United Kingdom}
\newcommand{\Eotvos}{E{\"o}tv{\"o}s Lor{\'a}nd University, 1053 Budapest, Hungary}
\newcommand{\FCULport}{Faculdade de Ci{\^e}ncias da Universidade de Lisboa - FCUL, 1749-016 Lisboa, Portugal}
\newcommand{\FederaldeAlfenas}{Universidade Federal de Alfenas, Po{\c{c}}os de Caldas - MG, 37715-400, Brazil}
\newcommand{\FederaldeGoias}{Universidade Federal de Goias, Goiania, GO 74690-900, Brazil}
\newcommand{\FederaldeSaoCarlos}{Universidade Federal de S{\~a}o Carlos, Araras - SP, 13604-900, Brazil}
\newcommand{\FederaldoABC}{Universidade Federal do ABC, Santo Andr{\'e} - SP, 09210-580, Brazil}
\newcommand{\FederaldoRio}{Universidade Federal do Rio de Janeiro,  Rio de Janeiro - RJ, 21941-901, Brazil}
\newcommand{\Fermi}{Fermi National Accelerator Laboratory, Batavia, IL 60510, USA}
\newcommand{\Ferrarauniv}{University of Ferrara, Ferrara, Italy}
\newcommand{\Florida}{University of Florida, Gainesville, FL 32611-8440, USA}
\newcommand{\Fluminense}{Fluminense Federal University, 9 Icara{\'\i} Niter{\'o}i - RJ, 24220-900, Brazil }
\newcommand{\Genova}{Universit{\`a} degli Studi di Genova, Genova, Italy}
\newcommand{\Georgian}{Georgian Technical University, Tbilisi, Georgia}
\newcommand{\GranSasso}{Gran Sasso Science Institute, L'Aquila, Italy}
\newcommand{\GranSassoLab}{Laboratori Nazionali del Gran Sasso, L'Aquila AQ, Italy}
\newcommand{\Granada}{University of Granada {\&} CAFPE, 18002 Granada, Spain}
\newcommand{\Grenoble}{University Grenoble Alpes, CNRS, Grenoble INP, LPSC-IN2P3, 38000 Grenoble, France}
\newcommand{\Guanajuato}{Universidad de Guanajuato, Guanajuato, C.P. 37000, Mexico}
\newcommand{\Harish}{Harish-Chandra Research Institute, Jhunsi, Allahabad 211 019, India}
\newcommand{\Harvard}{Harvard University, Cambridge, MA 02138, USA}
\newcommand{\Hawaii}{University of Hawaii, Honolulu, HI 96822, USA}
\newcommand{\Houston}{University of Houston, Houston, TX 77204, USA}
\newcommand{\Hyderabad}{University of  Hyderabad, Gachibowli, Hyderabad - 500 046, India}
\newcommand{\IFAE}{Institut de F{\'\i}sica d{\textquoteright}Altes Energies (IFAE){\textemdash}Barcelona Institute of Science and Technology (BIST), Barcelona, Spain}
\newcommand{\IFIC}{Instituto de F{\'\i}sica Corpuscular, CSIC and Universitat de Val{\`e}ncia, 46980 Paterna, Valencia, Spain}
\newcommand{\IGFAE}{Instituto Galego de Fisica de Altas Enerxias, A Coru{\~n}a, Spain}
\newcommand{\INFNBologna}{Istituto Nazionale di Fisica Nucleare Sezione di Bologna, 40127 Bologna BO, Italy}
\newcommand{\INFNCatania}{Istituto Nazionale di Fisica Nucleare Sezione di Catania, I-95123 Catania, Italy}
\newcommand{\INFNFerrara}{Istituto Nazionale di Fisica Nucleare Sezione di Ferrara, I-44122 Ferrara, Italy}
\newcommand{\INFNGenova}{Istituto Nazionale di Fisica Nucleare Sezione di Genova, 16146 Genova GE, Italy}
\newcommand{\INFNLecce}{Istituto Nazionale di Fisica Nucleare Sezione di Lecce, 73100 - Lecce, Italy}
\newcommand{\INFNMilanBicocca}{Istituto Nazionale di Fisica Nucleare Sezione di Milano Bicocca, 3 - I-20126 Milano, Italy}
\newcommand{\INFNMilano}{Istituto Nazionale di Fisica Nucleare Sezione di Milano, 20133 Milano, Italy}
\newcommand{\INFNNapoli}{Istituto Nazionale di Fisica Nucleare Sezione di Napoli, I-80126 Napoli, Italy}
\newcommand{\INFNPadova}{Istituto Nazionale di Fisica Nucleare Sezione di Padova, 35131 Padova, Italy}
\newcommand{\INFNPavia}{Istituto Nazionale di Fisica Nucleare Sezione di Pavia,  I-27100 Pavia, Italy}
\newcommand{\INFNRoma}{Istituto Nazionale di Fisica Nucleare Sezione di Roma, 00185 Roma RM}
\newcommand{\INFNSud}{Istituto Nazionale di Fisica Nucleare Laboratori Nazionali del Sud, 95123 Catania, Italy}
\newcommand{\INR}{Institute for Nuclear Research of the Russian Academy of Sciences, Moscow 117312, Russia}
\newcommand{\IPLyon}{Institut de Physique des 2 Infinis de Lyon, 69622 Villeurbanne, France}
\newcommand{\IPM}{Institute for Research in Fundamental Sciences, Tehran, Iran}
\newcommand{\ISTlisboa}{Instituto Superior T{\'e}cnico - IST, Universidade de Lisboa, Portugal}
\newcommand{\Idaho}{Idaho State University, Pocatello, ID 83209, USA}
\newcommand{\Illinoisinstitute}{Illinois Institute of Technology, Chicago, IL 60616, USA}
\newcommand{\Imperial}{Imperial College of Science Technology and Medicine, London SW7 2BZ, United Kingdom}
\newcommand{\IndGuwahati}{Indian Institute of Technology Guwahati, Guwahati, 781 039, India}
\newcommand{\IndHyderabad}{Indian Institute of Technology Hyderabad, Hyderabad, 502285, India}
\newcommand{\Indiana}{Indiana University, Bloomington, IN 47405, USA}
\newcommand{\Ingenieria}{Universidad Nacional de Ingenier{\'\i}a, Lima 25, Per{\'u}}
\newcommand{\Insubria }{University of Insubria, Via Ravasi, 2, 21100 Varese VA, Italy}
\newcommand{\Iowa}{University of Iowa, Iowa City, IA 52242, USA}
\newcommand{\IowaState}{Iowa State University, Ames, Iowa 50011, USA}
\newcommand{\Iwate}{Iwate University, Morioka, Iwate 020-8551, Japan}
\newcommand{\JINR}{Joint Institute for Nuclear Research, Dzhelepov Laboratory of Nuclear Problems 6 Joliot-Curie, Dubna, Moscow Region, 141980 RU }
\newcommand{\Jammu}{University of Jammu, Jammu-180006, India}
\newcommand{\Jawaharlal}{Jawaharlal Nehru University, New Delhi 110067, India}
\newcommand{\Jeonbuk}{Jeonbuk National University, Jeonrabuk-do 54896, South Korea}
\newcommand{\Jyvaskyla}{University of Jyvaskyla, FI-40014, Finland}
\newcommand{\KEK}{High Energy Accelerator Research Organization (KEK), Ibaraki, 305-0801, Japan}
\newcommand{\KISTI}{Korea Institute of Science and Technology Information, Daejeon, 34141, South Korea}
\newcommand{\KL}{K L University, Vaddeswaram, Andhra Pradesh 522502, India}
\newcommand{\Kansasstate}{Kansas State University, Manhattan, KS 66506, USA}
\newcommand{\Kavli}{Kavli Institute for the Physics and Mathematics of the Universe, Kashiwa, Chiba 277-8583, Japan}
\newcommand{\Kure}{National Institute of Technology, Kure College, Hiroshima, 737-8506, Japan}
\newcommand{\Kyiv}{Taras Shevchenko National University of Kyiv, 01601 Kyiv, Ukraine}
\newcommand{\LIP}{Laborat{\'o}rio de Instrumenta{\c{c}}{\~a}o e F{\'\i}sica Experimental de Part{\'\i}culas, 1649-003 Lisboa and 3004-516 Coimbra, Portugal}
\newcommand{\Lancaster}{Lancaster University, Lancaster LA1 4YB, United Kingdom}
\newcommand{\LawrenceBerkeley}{Lawrence Berkeley National Laboratory, Berkeley, CA 94720, USA}
\newcommand{\Liverpool}{University of Liverpool, L69 7ZE, Liverpool, United Kingdom}
\newcommand{\LosAlmos}{Los Alamos National Laboratory, Los Alamos, NM 87545, USA}
\newcommand{\Louisanastate}{Louisiana State University, Baton Rouge, LA 70803, USA}
\newcommand{\Lucknow}{University of Lucknow, Uttar Pradesh 226007, India}
\newcommand{\Madrid}{Madrid Autonoma University and IFT UAM/CSIC, 28049 Madrid, Spain}
\newcommand{\Mainz}{Johannes Gutenberg-Universit{\"a}t Mainz, 55122 Mainz, Germany}
\newcommand{\Manchester}{University of Manchester, Manchester M13 9PL, United Kingdom}
\newcommand{\Massinsttech}{Massachusetts Institute of Technology, Cambridge, MA 02139, USA}
\newcommand{\Maxplanck}{Max-Planck-Institut, Munich, 80805, Germany}
\newcommand{\Medellin}{University of Medell{\'\i}n, Medell{\'\i}n, 050026 Colombia }
\newcommand{\Michigan}{University of Michigan, Ann Arbor, MI 48109, USA}
\newcommand{\Michiganstate}{Michigan State University, East Lansing, MI 48824, USA}
\newcommand{\MilanoBicocca}{Universit{\`a} del Milano-Bicocca, 20126 Milano, Italy}
\newcommand{\MilanoUniv}{Universit{\`a} degli Studi di Milano, I-20133 Milano, Italy}
\newcommand{\Minnduluth}{University of Minnesota Duluth, Duluth, MN 55812, USA}
\newcommand{\Minntwin}{University of Minnesota Twin Cities, Minneapolis, MN 55455, USA}
\newcommand{\Mississippi}{University of Mississippi, University, MS 38677 USA}
\newcommand{\Newmexico}{University of New Mexico, Albuquerque, NM 87131, USA}
\newcommand{\Niewodniczanski}{H. Niewodnicza{\'n}ski Institute of Nuclear Physics, Polish Academy of Sciences, Cracow, Poland}
\newcommand{\Nikhef}{Nikhef National Institute of Subatomic Physics, 1098 XG Amsterdam, Netherlands}
\newcommand{\Northdakota}{University of North Dakota, Grand Forks, ND 58202-8357, USA}
\newcommand{\Northernillinois}{Northern Illinois University, DeKalb, IL 60115, USA}
\newcommand{\Northwestern}{Northwestern University, Evanston, Il 60208, USA}
\newcommand{\NotreDame}{University of Notre Dame, Notre Dame, IN 46556, USA}
\newcommand{\Occidental}{Occidental College, Los Angeles, CA  90041}
\newcommand{\Ohiostate}{Ohio State University, Columbus, OH 43210, USA}
\newcommand{\OregonState}{Oregon State University, Corvallis, OR 97331, USA}
\newcommand{\Oxford}{University of Oxford, Oxford, OX1 3RH, United Kingdom}
\newcommand{\PacificNorthwest}{Pacific Northwest National Laboratory, Richland, WA 99352, USA}
\newcommand{\Padova}{Universt{\`a} degli Studi di Padova, I-35131 Padova, Italy}
\newcommand{\Panjab}{Panjab University, Chandigarh, 160014 U.T., India}
\newcommand{\Parissaclay}{Universit{\'e} Paris-Saclay, CNRS/IN2P3, IJCLab, 91405 Orsay, France}
\newcommand{\Parisuniversite}{Universit{\'e} de Paris, CNRS, Astroparticule et Cosmologie, F-75006, Paris, France}
\newcommand{\Parma}{University of Parma,  43121 Parma PR, Italy}
\newcommand{\Pavia}{Universit{\`a} degli Studi di Pavia, 27100 Pavia PV, Italy}
\newcommand{\Penn}{University of Pennsylvania, Philadelphia, PA 19104, USA}
\newcommand{\PennState}{Pennsylvania State University, University Park, PA 16802, USA}
\newcommand{\PhysicalResearchLaboratory}{Physical Research Laboratory, Ahmedabad 380 009, India}
\newcommand{\Pisa}{Universit{\`a} di Pisa, I-56127 Pisa, Italy}
\newcommand{\Pitt}{University of Pittsburgh, Pittsburgh, PA 15260, USA}
\newcommand{\Pontificia}{Pontificia Universidad Cat{\'o}lica del Per{\'u}, Lima, Per{\'u}}
\newcommand{\PuertoRico}{University of Puerto Rico, Mayaguez 00681, Puerto Rico, USA}
\newcommand{\Punjab}{Punjab Agricultural University, Ludhiana 141004, India}
\newcommand{\QMUL}{Queen Mary University of London, London E1 4NS, United Kingdom }
\newcommand{\Radboud}{Radboud University, NL-6525 AJ Nijmegen, Netherlands}
\newcommand{\Rochester}{University of Rochester, Rochester, NY 14627, USA}
\newcommand{\Royalholloway}{Royal Holloway College London, TW20 0EX, United Kingdom}
\newcommand{\Rutgers}{Rutgers University, Piscataway, NJ, 08854, USA}
\newcommand{\Rutherford}{STFC Rutherford Appleton Laboratory, Didcot OX11 0QX, United Kingdom}
\newcommand{\SLAC}{SLAC National Accelerator Laboratory, Menlo Park, CA 94025, USA}
\newcommand{\SURF}{Sanford Underground Research Facility, Lead, SD, 57754, USA}
\newcommand{\Salento}{Universit{\`a} del Salento, 73100 Lecce, Italy}
\newcommand{\Sanjosestate}{San Jose State University, San Jos{\'e}, CA 95192-0106, USA}
\newcommand{\Sapienza}{Sapienza University of Rome, 00185 Roma RM, Italy}
\newcommand{\SergioArboleda}{Universidad Sergio Arboleda, 11022 Bogot{\'a}, Colombia}
\newcommand{\Sheffield}{University of Sheffield, Sheffield S3 7RH, United Kingdom}
\newcommand{\SouthDakotaSchool}{South Dakota School of Mines and Technology, Rapid City, SD 57701, USA}
\newcommand{\SouthDakotaState}{South Dakota State University, Brookings, SD 57007, USA}
\newcommand{\Southcarolina}{University of South Carolina, Columbia, SC 29208, USA}
\newcommand{\SouthernMethodist}{Southern Methodist University, Dallas, TX 75275, USA}
\newcommand{\StonyBrook}{Stony Brook University, SUNY, Stony Brook, NY 11794, USA}
\newcommand{\Sunyatsen}{Sun Yat-Sen University, Guangzhou, 510275}
\newcommand{\Sussex}{University of Sussex, Brighton, BN1 9RH, United Kingdom}
\newcommand{\Syracuse}{Syracuse University, Syracuse, NY 13244, USA}
\newcommand{\Tecnologica }{Universidade Tecnol{\'o}gica Federal do Paran{\'a}, Curitiba, Brazil}
\newcommand{\TexasAMcollege}{Texas A{\&}M University, College Station, Texas 77840}
\newcommand{\TexasAMcorpuscristi}{Texas A{\&}M University - Corpus Christi, Corpus Christi, TX 78412, USA}
\newcommand{\TexasArlington}{University of Texas at Arlington, Arlington, TX 76019, USA}
\newcommand{\Texasaustin}{University of Texas at Austin, Austin, TX 78712, USA}
\newcommand{\Toronto}{University of Toronto, Toronto, Ontario M5S 1A1, Canada}
\newcommand{\Tufts}{Tufts University, Medford, MA 02155, USA}
\newcommand{\UNIST}{Ulsan National Institute of Science and Technology, Ulsan 689-798, South Korea}
\newcommand{\Unifesp}{Universidade Federal de S{\~a}o Paulo, 09913-030, S{\~a}o Paulo, Brazil}
\newcommand{\UniversityCollegeLondon}{University College London, London, WC1E 6BT, United Kingdom}
\newcommand{\ValleyCity}{Valley City State University, Valley City, ND 58072, USA}
\newcommand{\VariableEnergy}{Variable Energy Cyclotron Centre, 700 064 West Bengal, India}
\newcommand{\VirginiaTech}{Virginia Tech, Blacksburg, VA 24060, USA}
\newcommand{\Warsaw}{University of Warsaw, 02-093 Warsaw, Poland}
\newcommand{\Warwick}{University of Warwick, Coventry CV4 7AL, United Kingdom}
\newcommand{\Wellesley}{Wellesley College, Wellesley, MA 02481, USA}
\newcommand{\Wichita}{Wichita State University, Wichita, KS 67260, USA}
\newcommand{\WilliamMary}{William and Mary, Williamsburg, VA 23187, USA}
\newcommand{\Wisconsin}{University of Wisconsin Madison, Madison, WI 53706, USA}
\newcommand{\Yale}{Yale University, New Haven, CT 06520, USA}
\newcommand{\Yerevan}{Yerevan Institute for Theoretical Physics and Modeling, Yerevan 0036, Armenia}
\newcommand{\York}{York University, Toronto M3J 1P3, Canada}
\newcommand{\napoli}{Universit{\`a} degli Studi di Napoli Federico II , 80138 Napoli NA, Italy}
\affiliation{\Abilene}
\affiliation{\Albanysuny}
\affiliation{\Amsterdam}
\affiliation{\Antalya}
\affiliation{\Antananarivo}
\affiliation{\AntonioNarino}
\affiliation{\Argonne}
\affiliation{\Arizona}
\affiliation{\Asuncion}
\affiliation{\Athens}
\affiliation{\Atlantico}
\affiliation{\Augustana}
\affiliation{\Basel}
\affiliation{\Bern}
\affiliation{\Beykent}
\affiliation{\Birmingham}
\affiliation{\BolognaUniversity}
\affiliation{\Boston}
\affiliation{\Bristol}
\affiliation{\Brookhaven}
\affiliation{\Bucharest}
\affiliation{\CBPF}
\affiliation{\CEASaclay}
\affiliation{\CERN}
\affiliation{\CIEMAT}
\affiliation{\CUSB}
\affiliation{\CalBerkeley}
\affiliation{\CalDavis}
\affiliation{\CalIrvine}
\affiliation{\CalLosangeles}
\affiliation{\CalRiverside}
\affiliation{\CalSantabarbara}
\affiliation{\Caltech}
\affiliation{\Cambridge}
\affiliation{\Campinas}
\affiliation{\CataniaUniversitadi}
\affiliation{\Catolica}
\affiliation{\Charles}
\affiliation{\Chicago}
\affiliation{\ChungAng}
\affiliation{\Cincinnati}
\affiliation{\Cinvestav}
\affiliation{\Colima}
\affiliation{\ColoradoBoulder}
\affiliation{\ColoradoState}
\affiliation{\Columbia}
\affiliation{\Cti}
\affiliation{\CzechAcademyofSciences}
\affiliation{\CzechTechnical}
\affiliation{\DakotaState}
\affiliation{\Dallas}
\affiliation{\DannecyleVieux}
\affiliation{\Daresbury}
\affiliation{\Drexel}
\affiliation{\Duke}
\affiliation{\Durham}
\affiliation{\EIA}
\affiliation{\ETH}
\affiliation{\Edinburgh}
\affiliation{\Eotvos}
\affiliation{\FCULport}
\affiliation{\FederaldeAlfenas}
\affiliation{\FederaldeGoias}
\affiliation{\FederaldeSaoCarlos}
\affiliation{\FederaldoABC}
\affiliation{\FederaldoRio}
\affiliation{\Fermi}
\affiliation{\Ferrarauniv}
\affiliation{\Florida}
\affiliation{\Fluminense}
\affiliation{\Genova}
\affiliation{\Georgian}
\affiliation{\GranSasso}
\affiliation{\GranSassoLab}
\affiliation{\Granada}
\affiliation{\Grenoble}
\affiliation{\Guanajuato}
\affiliation{\Harish}
\affiliation{\Harvard}
\affiliation{\Hawaii}
\affiliation{\Houston}
\affiliation{\Hyderabad}
\affiliation{\IFAE}
\affiliation{\IFIC}
\affiliation{\IGFAE}
\affiliation{\INFNBologna}
\affiliation{\INFNCatania}
\affiliation{\INFNFerrara}
\affiliation{\INFNGenova}
\affiliation{\INFNLecce}
\affiliation{\INFNMilanBicocca}
\affiliation{\INFNMilano}
\affiliation{\INFNNapoli}
\affiliation{\INFNPadova}
\affiliation{\INFNPavia}
\affiliation{\INFNRoma}
\affiliation{\INFNSud}
\affiliation{\INR}
\affiliation{\IPLyon}
\affiliation{\IPM}
\affiliation{\ISTlisboa}
\affiliation{\Idaho}
\affiliation{\Illinoisinstitute}
\affiliation{\Imperial}
\affiliation{\IndGuwahati}
\affiliation{\IndHyderabad}
\affiliation{\Indiana}
\affiliation{\Ingenieria}
\affiliation{\Insubria }
\affiliation{\Iowa}
\affiliation{\IowaState}
\affiliation{\Iwate}
\affiliation{\JINR}
\affiliation{\Jammu}
\affiliation{\Jawaharlal}
\affiliation{\Jeonbuk}
\affiliation{\Jyvaskyla}
\affiliation{\KEK}
\affiliation{\KISTI}
\affiliation{\KL}
\affiliation{\Kansasstate}
\affiliation{\Kavli}
\affiliation{\Kure}
\affiliation{\Kyiv}
\affiliation{\LIP}
\affiliation{\Lancaster}
\affiliation{\LawrenceBerkeley}
\affiliation{\Liverpool}
\affiliation{\LosAlmos}
\affiliation{\Louisanastate}
\affiliation{\Lucknow}
\affiliation{\Madrid}
\affiliation{\Mainz}
\affiliation{\Manchester}
\affiliation{\Massinsttech}
\affiliation{\Maxplanck}
\affiliation{\Medellin}
\affiliation{\Michigan}
\affiliation{\Michiganstate}
\affiliation{\MilanoBicocca}
\affiliation{\MilanoUniv}
\affiliation{\Minnduluth}
\affiliation{\Minntwin}
\affiliation{\Mississippi}
\affiliation{\napoli}
\affiliation{\Newmexico}
\affiliation{\Niewodniczanski}
\affiliation{\Nikhef}
\affiliation{\Northdakota}
\affiliation{\Northernillinois}
\affiliation{\Northwestern}
\affiliation{\NotreDame}
\affiliation{\Occidental}
\affiliation{\Ohiostate}
\affiliation{\OregonState}
\affiliation{\Oxford}
\affiliation{\PacificNorthwest}
\affiliation{\Padova}
\affiliation{\Panjab}
\affiliation{\Parissaclay}
\affiliation{\Parisuniversite}
\affiliation{\Parma}
\affiliation{\Pavia}
\affiliation{\Penn}
\affiliation{\PennState}
\affiliation{\PhysicalResearchLaboratory}
\affiliation{\Pisa}
\affiliation{\Pitt}
\affiliation{\Pontificia}
\affiliation{\PuertoRico}
\affiliation{\Punjab}
\affiliation{\QMUL}
\affiliation{\Radboud}
\affiliation{\Rochester}
\affiliation{\Royalholloway}
\affiliation{\Rutgers}
\affiliation{\Rutherford}
\affiliation{\SLAC}
\affiliation{\SURF}
\affiliation{\Salento}
\affiliation{\Sanjosestate}
\affiliation{\Sapienza}
\affiliation{\SergioArboleda}
\affiliation{\Sheffield}
\affiliation{\SouthDakotaSchool}
\affiliation{\SouthDakotaState}
\affiliation{\Southcarolina}
\affiliation{\SouthernMethodist}
\affiliation{\StonyBrook}
\affiliation{\Sunyatsen}
\affiliation{\Sussex}
\affiliation{\Syracuse}
\affiliation{\Tecnologica }
\affiliation{\TexasAMcollege}
\affiliation{\TexasAMcorpuscristi}
\affiliation{\TexasArlington}
\affiliation{\Texasaustin}
\affiliation{\Toronto}
\affiliation{\Tufts}
\affiliation{\UNIST}
\affiliation{\Unifesp}
\affiliation{\UniversityCollegeLondon}
\affiliation{\ValleyCity}
\affiliation{\VariableEnergy}
\affiliation{\VirginiaTech}
\affiliation{\Warsaw}
\affiliation{\Warwick}
\affiliation{\Wellesley}
\affiliation{\Wichita}
\affiliation{\WilliamMary}
\affiliation{\Wisconsin}
\affiliation{\Yale}
\affiliation{\Yerevan}
\affiliation{\York}

\author{A.~Abed Abud} \affiliation{\Liverpool}\affiliation{\CERN}
\author{B.~Abi} \affiliation{\Oxford}
\author{R.~Acciarri} \affiliation{\Fermi}
\author{M.~A.~Acero} \affiliation{\Atlantico}
\author{M.~R.~Adames} \affiliation{\Tecnologica }
\author{G.~Adamov} \affiliation{\Georgian}
\author{M.~Adamowski} \affiliation{\Fermi}
\author{D.~Adams} \affiliation{\Brookhaven}
\author{M.~Adinolfi} \affiliation{\Bristol}
\author{C.~Adriano} \affiliation{\Campinas}
% ...
\collaboration{The DUNE Collaboration}
\noaffiliation











\maketitle

\snowmass


\flushbottom

\pagebreak

\section*{Executive Summary}
\label{sec:summary}

This document presents the concept and physics case for a magnetized gaseous argon-based detector system (ND-GAr) for the Deep Underground Neutrino Experiment (DUNE) Near Detector.  This detector system is required in order for DUNE to reach its full physics potential in the measurement of CP violation and in delivering precision measurements of oscillation parameters.

The ND-GAr concept is based on a central high-pressure gaseous argon time projection chamber (HPgTPC); the HPgTPC is surrounded by a calorimeter, with both situated in a 0.5~T magnetic field generated by superconducting coils. A muon system is integrated with the magnet return yoke. The baseline concept of the HPgTPC at the heart of ND-GAr is based closely on the design of the ALICE TPC~\cite{Alme:2010ke}. An opportunity exists to reconsider this design, given the 
timeline to construct the full DUNE. This paper focuses on the baseline design in order to present physics motivations and performance. Future R\&D lines aimed at optimizing this design are discussed as well.  

The DUNE CP violation measurement requires complementary detectors at the near site: an upstream modular non-magnetized liquid argon TPC (ND-LAr), a magnetized tracker containing a pressurized gaseous argon TPC (ND-GAr), and a large magnetized beam monitor (SAND). In the early years of DUNE running, a smaller and simpler detector than ND-GAr can be placed downstream of ND-LAr to measure the momentum and charge sign of forward-going muons exiting ND-LAr. However, after approximately 200 kt-MW-yrs of exposure, the full ND-GAr detector is critical for DUNE to reach 5$\sigma$ sensitivity to CP violation over most of the possible values of $\delta_{CP}$, the Standard Model parameter that controls CP violation in neutrinos. 

In addition to its critical role in the long-baseline oscillation program, ND-GAr will extend the overall physics program of DUNE. This detector will collect a significant number of neutrino interactions ($\sim$1.6 M/yr) on its one-ton gaseous argon target, enabling a variety of neutrino interaction cross section measurements capable of detecting the resulting charged particles with lower energies than achievable in the far or near liquid argon TPCs, and with enhanced particle identification performance relative to the LArTPCs. These capabilities enable better constraints of systematic uncertainties for both oscillation and cross section analyses.  Another unique advantage of ND-GAr is its flexibility to use various gas mixtures as interaction targets, such as potentially using a hydrogen-rich gas mixture to probe more fundamental neutrino-hydrogen interactions, which can also better constrain understanding of neutrino-argon interactions. 

The LBNF high-intensity proton beam will provide a large flux of neutrinos that is sampled by ND-GAr, enabling DUNE to discover new particles and search for new interactions and symmetries beyond those predicted in the Standard Model. In particular, ND-GAr can search for neutrino tridents, heavy neutral leptons, light dark matter, heavy axions, and anomalous tau neutrinos from short-baseline mixing with sterile neutrinos.

The baseline design of ND-GAr takes advantage of much work done on the ALICE TPC and by the CALICE calorimeter collaboration, but it is necessary to adapt these to the DUNE near detector environment.  A focused R\&D program is presented, aimed at optimization of the detector design along with exploration of timely new technological developments that offer a window of opportunity for extending the capabilities of DUNE.
\newpage

\tableofcontents

\newpage
\section{Introduction}
\label{sec:intro}
The Deep Underground Neutrino Experiment (DUNE) is a next-generation international particle physics experiment seeking to answer fundamental questions about the neutrino.  
It will use a new high-intensity neutrino beam that will be generated at the U.S. Department of Energy's Fermi National Accelerator Laboratory (Fermilab). 
The experiment will consist of a far detector (FD) located approximately 1.5~km underground at the Sanford Underground Research Facility (SURF) in Lead, South Dakota, at a distance of 1285~km from Fermilab, and a near detector (ND) that will be located on the Fermilab site in Illinois.  The FD will consist of a large, modular liquid-argon time projection chamber (LArTPC) with a fiducial mass of roughly 40~kt (total mass of 68 kton).

\begin{figure}[ht]
\centering
\includegraphics[width=0.95\textwidth]{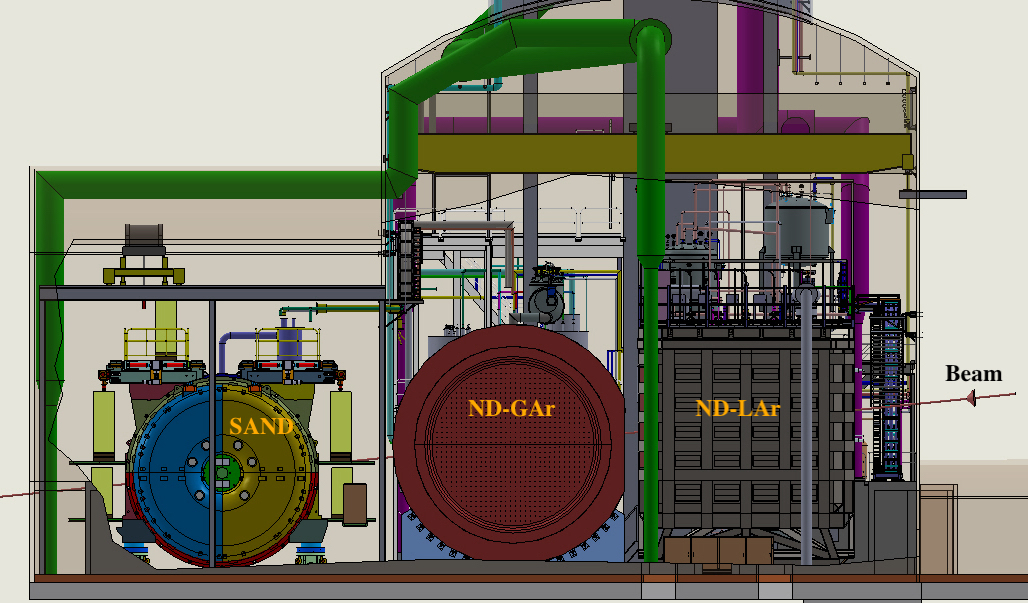}
\caption[DUNE ND Hall]{Profile view of the DUNE ND hall and the reference DUNE ND detectors foreseen to operate in Phase~II.  The neutrino beam enters from the right.}
\label{fig:ndlayout}
\end{figure}

The reference ND will be located approximately 574~m from DUNE's
neutrino target, which is the starting point for the Long-Baseline Neutrino Facility (LBNF) beam. The conceptual design for the DUNE ND is described in detail in Ref.~\cite{DUNE:2021tad}. 
The reference ND design consists of several different components shown in Fig.~\ref{fig:ndlayout}: an upstream modular non-magnetized LArTPC (ND-LAr), a magnetized tracker containing a pressurized gaseous-argon time projection chamber (ND-GAr), and a large, magnetized tracking spectrometer (SAND).  SAND will remain fixed on the beam axis, while ND-LAr and ND-GAr will move transverse to the beam to collect data at various off-axis positions, providing different neutrino energy spectra by the PRISM concept~\cite{bhadra2014letter}.  In a conventional neutrino beam, the peak energy of the neutrino spectrum decreases and the size of the high energy tail is reduced when the detection angle relative to the beam axis is increased.  For this reason the DUNE-PRISM measurement program consists of moving ND-LAr and ND-GAr laterally, transverse to the beam's direction through the underground ND hall, over an off-axis angle range of 0$^\circ$ to 3.2$^\circ$  (a distance of $\sim$31~m).  These measurements will allow for a data-driven determination of the relationship between true and reconstructed energy and will provide data samples that can be combined to produce a flux at the ND that is very similar to the expected oscillated flux at the FD. 

While ND-GAr has a long history in DUNE and is part of the reference design, 
the near detector complex has been organized into a phased program, in which Phase~I contains a minimal tracking detector downstream of ND-LAr, and Phase~II includes the upgrade to ND-GAr that is necessary to achieve the DUNE physics goals. This white paper presents the arguments for why ND-GAr is needed in the long term and also includes the R\&D program needed to get there.

\subsection*{Detector Concept}

In order to constrain neutrino interaction model uncertainties, it is essential that the near detector contain an active argon medium to measure neutrino interactions on the same target nucleus as the far detector.  The non-magnetized ND-LAr can measure neutrino interactions on argon with a similar detection principle as the far detector, but only muons below 1~GeV/$c$ will be fully contained in the detector. The FD has better acceptance due to its larger size, so the ND needs access to uncontained muons to match the types of events that will be seen in the FD.  Thus an additional magnetized detector is required in addition to ND-LAr to measure the momentum and charge-sign of the exiting muons.  

\begin{figure}[ht]
\centering
\includegraphics[width=\textwidth]{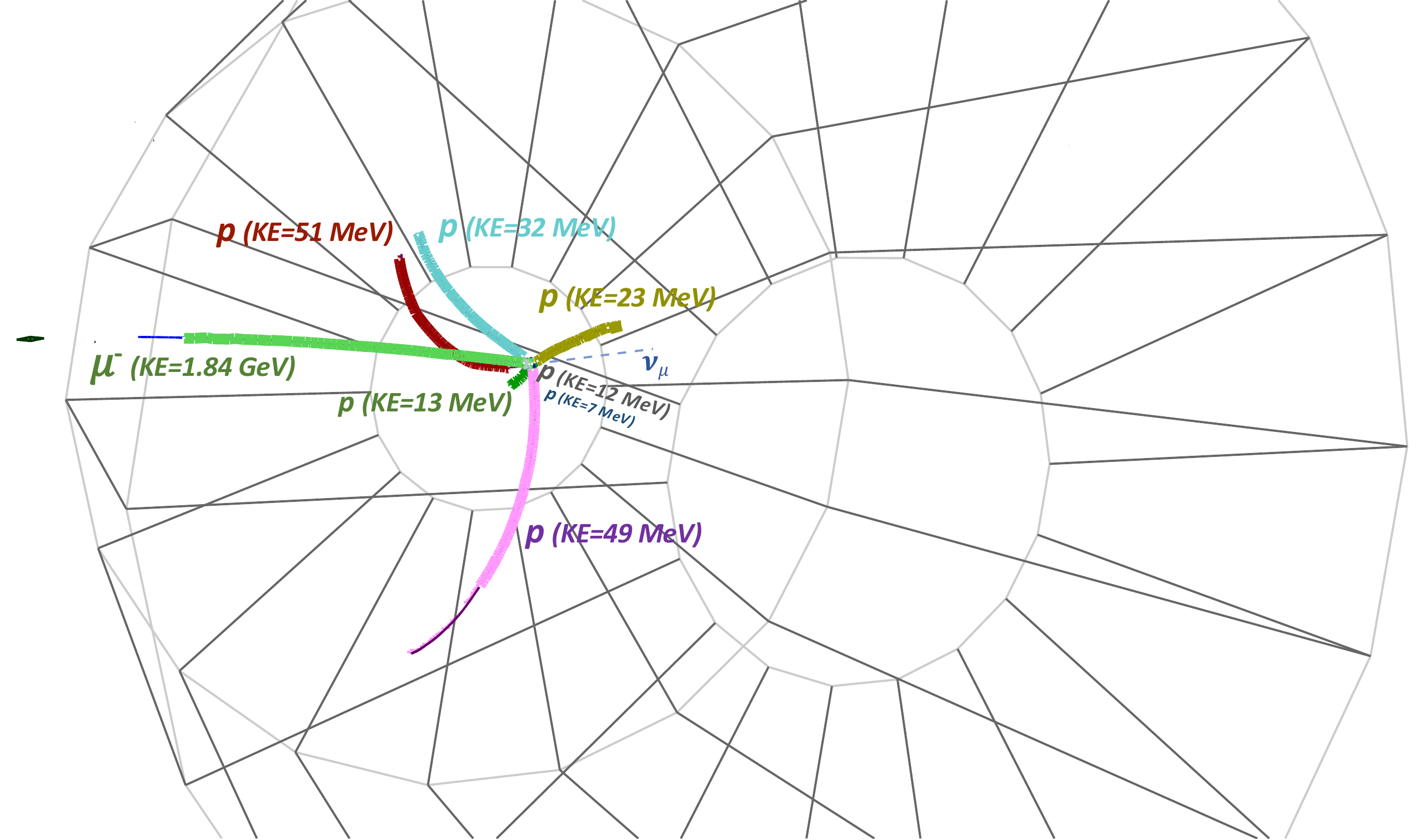}
\caption[GArSoft simulated and reconstructed event with low energy protons]{A charged current $\nu_\mu$ event with seven low energy protons (kinetic energies ranging from 7 to 51~MeV) in a 10~atm argon-based TPC, simulated and reconstructed with the GArSoft software suite. The detector has a radius of approximately 2.5 m and a width of 5 m.   Charged particle trajectories for both simulated (thin lines) and reconstructed (thick lines) tracks are shown. The $\mu^{-}$ (green reconstructed track) enters the ECAL, while the protons are contained within the TPC. The reconstruction algorithm finds all eight tracks, although only six are visible by eye in this view.
}
\label{fig:garsoft-evd-manyp}
\end{figure}

\begin{figure}[ht]
\centering
\includegraphics[width=0.75\textwidth]{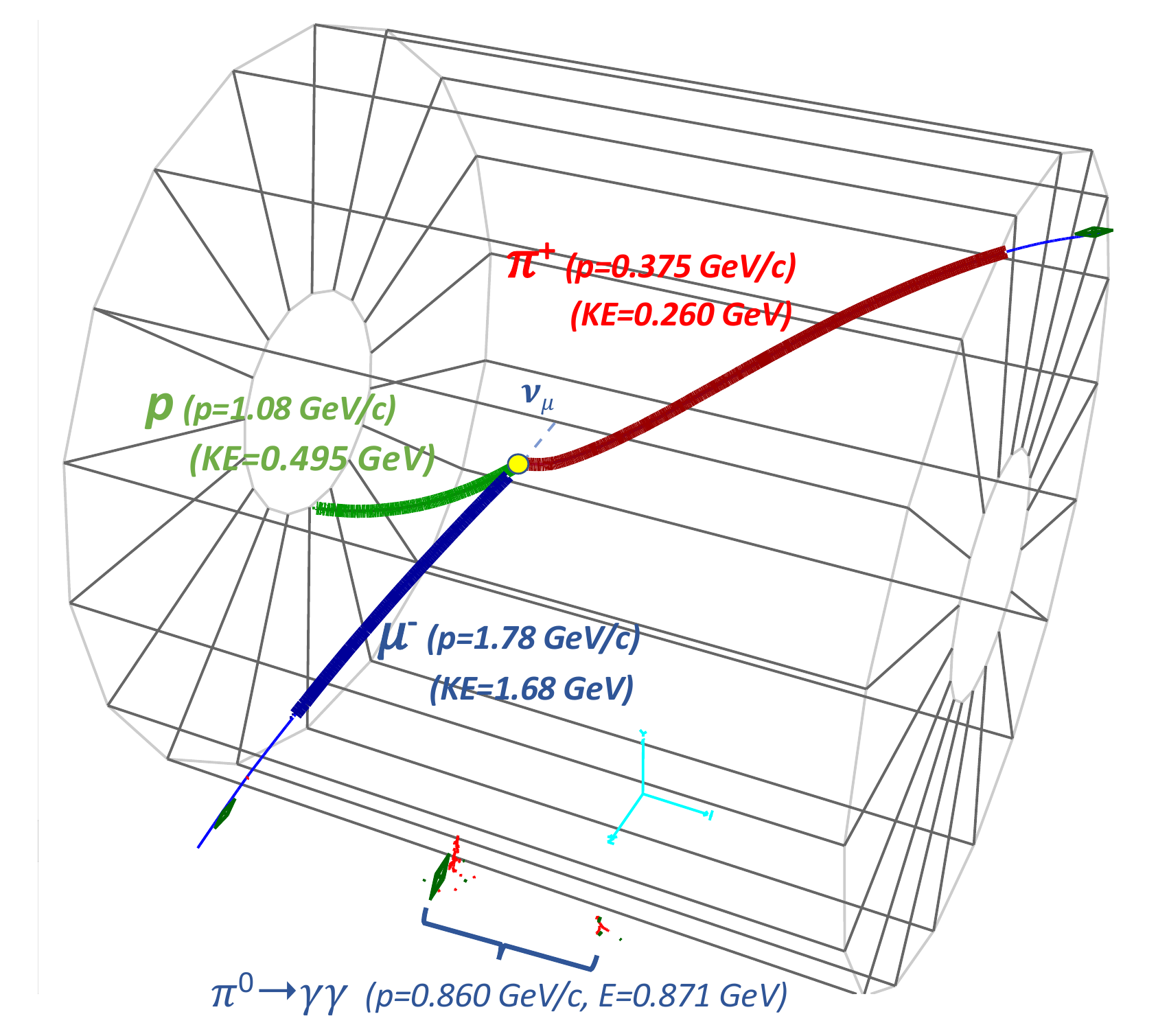}
\caption[GArSoft simulated and reconstructed event]{A charged current $\nu_\mu$ event with two pions in a 10~atm argon-based TPC, simulated and reconstructed with the GArSoft software suite. The detector has a radius of approximately 2.5 m and a width of 5 m.   The annotations are from Monte Carlo truth. Charged particle trajectories for both simulated (thin lines) and reconstructed (thick lines) tracks are shown, as well as reconstructed clusters in the ECAL (shown as green polyhedra). The $\pi^{+}$ decays in flight, and the resulting muon enters the ECAL. Also seen in the ECAL are reconstructed clusters from the decay of a final state $\pi^{0}$ created in the neutrino interaction (the two $\gamma$'s from the $\pi^{0}$ decay do not produce ionization tracks in the TPC, and so are only detected by the ECAL), and a cluster from the outgoing $\mu^{-}$ in the interaction. 
}
\label{fig:garsoft-evd}
\end{figure}

\begin{figure}[ht]
\centering
\includegraphics[width=0.5\textwidth]{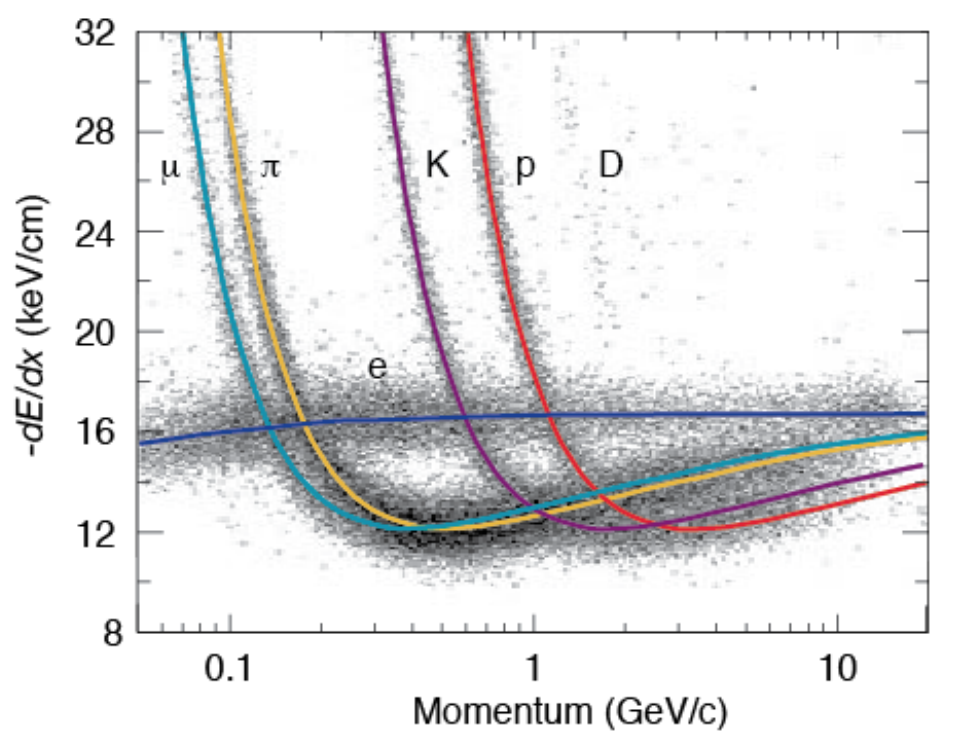}
\caption[PEP-4 TPC dE/dx]{$dE/dx$-based particle identification in the TPC of the PEP-4 detector at SLAC~\cite{Grupen:1999by}.  This TPC used a gas mixture of 80:20 Ar-CH$_4$, operated at 8.5~atm~\cite{TPCTwoGamma:1982efn}.}
\label{fig:pep4-dedx}
\end{figure}

A gaseous argon-based time projection chamber (gaseous-argon TPC, or GArTPC) can reconstruct pions and protons with lower detection thresholds than a liquid-argon TPC (LArTPC). For comparison, a proton must have a kinetic energy of 46~MeV in order to make a 2-cm track in liquid argon, but only it only needs 4~MeV to make the same length track in 10~atm of gaseous argon. Proton tracks have been successfully reconstructed in the MicroBooNE LArTPC down to kinetic energies of approximately 46~MeV (momentum of 300~MeV/$c$)~\cite{MicroBooNE:2020akw}. The GArTPC is also less susceptible to confusion of primary and secondary interactions, since secondary interactions occur infrequently in the lower density gas detector ($\lambda \sim$90~m). If the TPC is inside a magnet, it can better distinguish neutrinos and anti-neutrinos and can determine the momenta of particles that range out of the detector. It can also measure neutrino interactions over all directions, unlike the ND-LAr, which loses acceptance at high angles with respect to the beam direction. Therefore, a gaseous-argon TPC provides a valuable and complementary sample of argon interactions to better understand neutrino-argon interactions.  An example of a simulated and track-reconstructed neutrino event in a gaseous-argon TPC is shown in Fig.~\ref{fig:garsoft-evd}. To ensure adequate statistics of more than 1 million events per year in a reasonably sized detector, this gaseous-argon TPC will need to operate at approximately 10~atm.  Using a gaseous TPC as the primary tracking element in the spectrometer allows for excellent particle momentum resolution and particle identification (PID). An example of the $dE/dx$-based PID capabilities achieved in the high-pressure TPC in the PEP-4 detector facility~\cite{TPCTwoGamma:1982efn} at SLAC is shown in Fig.~\ref{fig:pep4-dedx}. ND-GAr, at a similar or slightly higher pressure and larger sampling volume, is expected to achieve similar or better identification capabilities.

To detect neutrons and photons (mainly from neutral pions), the TPC must be surrounded by a calorimeter, though this might be thinner on the upstream side to keep acceptance high for tracks entering from ND-LAr.  A muon system outside the magnet is needed to help separate muons and charged pions.  Putting all of these general requirements together leads to the complete ND-GAr concept, with a central high-pressure gaseous-argon TPC (HPgTPC) surrounded by a calorimeter (ECAL) and a ~0.5 T magnetic field, and a muon tagger system.  More details of the current detector concept can be found in Section~\ref{sec:det}.


\section{Physics Motivation}
\label{sec:physicsmotivation}

The full physics program of DUNE is described in Refs.~\cite{DUNE:2015lol, Abi:2018dnh, DUNE:2020ypp}. The near detector for DUNE Phase I is planned to consist of ND-LAr, a Temporary Muon Spectrometer (TMS), which measures the charge and momentum of muons that exit ND-LAr but is not designed to perform additional precision physics measurements, and SAND. This configuration is shown to be sufficient for the early DUNE physics milestones, including determination of the neutrino mass ordering and sensitivity to maximal CP violation. To reach the precision physics goals, including sensitivity to CP violation over a broad range of values of $\delta_{CP}$, systematic uncertainties must be constrained to a level well beyond what is achieved by current experiments, and this requires a more capable near detector.  Even if CP violation is found to be large and near maximal, precise measurements will require ND-GAr.


 In addition to measuring the momentum and sign of charged particles exiting ND-LAr,  ND-GAr provides a magnetized gaseous argon target within ND-GAr, which will extend charged-particle measurement capabilities.
 It will also greatly extend the particle ID (PID) performance, particularly for proton-pion separation. These capabilities enable precision measurements of neutrino interaction cross sections to further constrain systematic uncertainties for the long-baseline oscillation analysis. Since the target nucleus in ND-GAr is the same as in the near and far LArTPCs, this information helps constrain interaction models by minimizing contributions from nuclear physics effects. The ND-GAr detector, with its full solid-angle acceptance, will be used to measure ratios of inclusive, semi-exclusive, and exclusive cross sections as functions of neutrino energy, where the flux cancels in the ratio. 
 Cross sections for different particle species multiplicities (pions, protons, kaons, etc.) will help constrain interaction models used in the near and far liquid argon detectors. ND-GAr will also replace the TMS in its capacity as a muon spectrometer for events originating in ND-LAr.

In addition to its critical role in the DUNE long-baseline oscillation program, ND-GAr will collect a significant number of neutrino interactions ($\sim$1.6 M/yr) on its gaseous argon target, enabling a broad stand-alone physics program of neutrino interaction cross section measurements including the study of exotic physics channels, as well as Beyond Standard Model searches. 

\subsection{Oscillation Physics}

The ND-GAr detector provides lower threshold measurements than ND-LAr can provide, making it possible to pin down the interaction model and associated systematic uncertainties needed for the oscillation program.   In particular, ND-GAr measures hadron species,  multiplicity, and momenta both to very low threshold and also at energies where hadrons would interact in the ND-LAr or exit the detector volume, thus preventing particle identification in ND-LAr.  This complementary advantage allows the oscillation analysis strategy to constrain interaction processes by outgoing particle content, which correlates strongly with momentum transfer, intranuclear reinteraction processes, and missing energy from neutrons that affect the neutrino energy estimate. Since these interaction processes are not strongly correlated with flux, ND-GAr will be able to disentangle these uncertainties.

\begin{figure}[ht]
\centering
\includegraphics[width=0.495\textwidth]{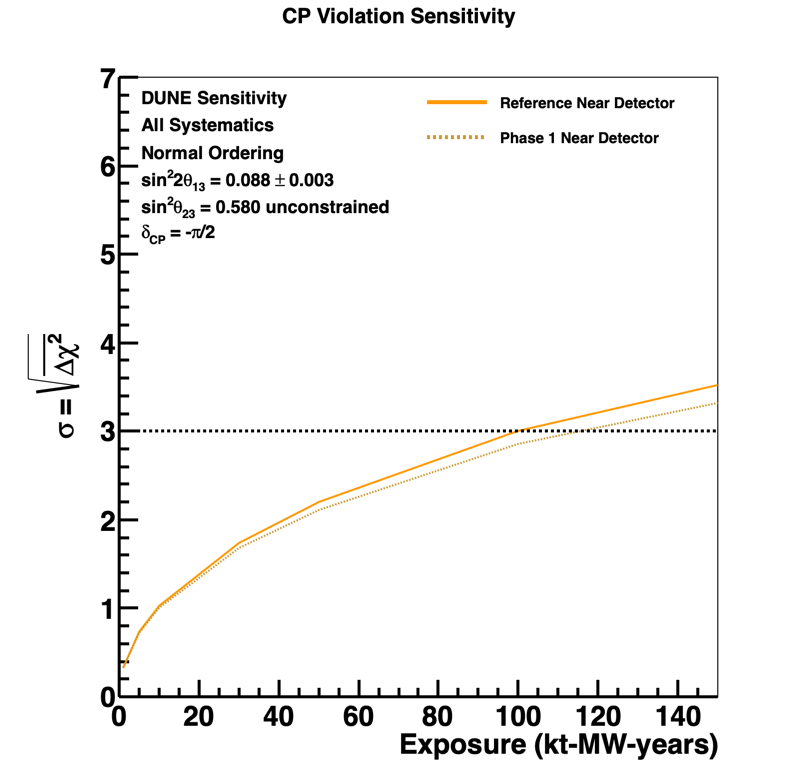}
\includegraphics[width=0.495\textwidth]{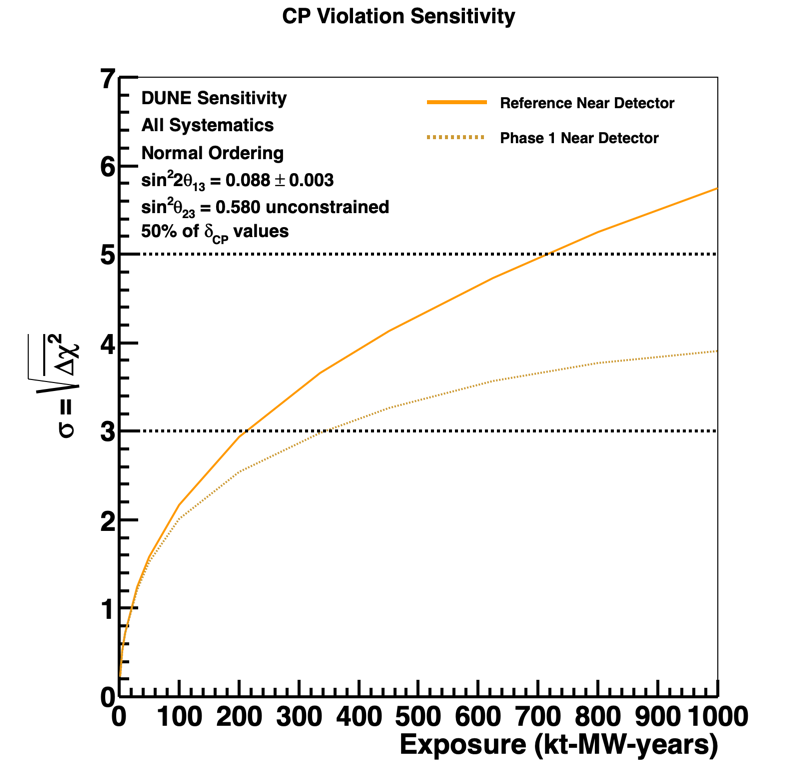}
\caption[Impact of full ND-GAr on CP violation sensitivity]{Impact of the Phase~I and reference near detector configurations on CP violation sensitivity, shown for maximal CP violation (left) and for 50\% coverage of $\delta_{CP}$ space.}
\label{fig:cpsens}
\end{figure}

Figure~\ref{fig:cpsens} shows the early running (left) and ultimate (right) impact of the ND-GAr detector on the long-baseline oscillation sensitivity to CP violation. The Phase I sensitivity is evaluated from a case study where data based on the NuWro generator~\cite{Golan:2012wx,Golan2012nuwro} are fit with the nominal GENIE model, resulting in a bias to oscillation parameters that can be mitigated by adding data from ND-GAr. As seen in the left panel, if CP violation is maximal, there is not much difference between the exposure required to achieve sensitivity with $3\sigma$ significance using TMS (Phase~I Near Detector) and ND-GAr (Reference Near Detector~\cite{Abi:2020qib, DUNE:2021tad}). Maximal CP violation is the most favorable case for DUNE's sensitivity, but a more realistic assessment of sensitivity is given by looking at the full range of possible values of $\delta_{CP}$ (or some fraction of that range of values). The right panel shows DUNE's expected sensitivity to 50\% of all $\delta_{CP}$ values, for the same detector configurations. In this case, the experiment with TMS  
may possibly never reach the $5\sigma$ significance level even with very long exposures. The full ND-GAr detector is ultimately needed to achieve DUNE's scientific goal for CP violation.  This need becomes more critical after DUNE reaches an exposure of approximately 200 kt-MW-yrs.


\subsection{Cross Section Physics}



Precision measurements of neutrino interaction cross sections are critical to the oscillation program, but are also interesting in and of themselves. In liquid argon detectors, secondary interactions are a significant effect~\cite{Friedland:2018vry}, since the hadronic interaction length of pions is $\lambda \sim$1~m in LAr. By contrast, in 10~atm of gaseous argon, $\lambda \sim$90~m.  The relative lack of secondary interactions for particles formed in neutrino interactions in the gaseous argon fiducial volume will provide event samples that are less dependent on understanding of detector response and models of secondary interactions. 
Since the target nucleus in ND-GAr is the same as that in the ND-LAr and far detectors, this information feeds directly into the interaction model constraints for the oscillation analysis, but without complicating nuclear physics concerns. These measurements will also be useful inputs for tuning the interaction models to better reflect what is seen in data, as the agreement among models and existing world data is not great.

ND-GAr, with its full solid-angle acceptance and MeV-level proton tracking threshold, is ideal for performing exclusive final-state measurements. In the exclusive channels, $\nu+\textrm{A}\rightarrow\ell+\textrm{hadrons}+\textrm{A}'$, where the final-state hadrons are measured, the details of the intranuclear dynamics of the interactions can be precisely extracted by exploring momentum conservation in the transverse plane to the neutrino direction, regardless of the unknown neutrino energy. This technique of measuring Transverse Kinematic Imbalance (TKI) in neutrino interactions~\cite{Lu:2015hea,Lu:2015tcr,Furmanski:2016wqo,Lu:2019nmf} has been successfully applied in MINERvA~\cite{MINERvA:2018hba,MINERvA:2019ope,MINERvA:2020anu} and T2K~\cite{T2K:2018rnz,T2K:2021naz}. With the state-of-the-art tracking resolution, ND-GAr will use TKI to further explore neutrino-argon interactions with unprecedented precision~\cite{DUNE:2021tad}.

The predicted event yields for one year of neutrino running are shown in Table~\ref{tab:Stats}, assuming the DUNE flux~\cite{DUNE:2020ypp}, 
cross-sections from GENIE~\cite{Dytman:2015xaa, Andreopoulos:2009rq} 2.12.2, and a fiducial mass of 1~t ($\simeq$ 55\% of the active mass).  The lower portion of the table highlights the expected yields for some of the exclusive final states for which ND-GAr will be able to make precise measurements. The baseline gas in ND-GAr is assumed to be 10~atm of an argon gas admixture of 90\% Ar and 10\% CH$_4$, which results in $\simeq 97\%$ of interactions occurring on Ar nuclei. 
%
%

\begin{table}[!ht]
    \centering
    \caption{Number of events in the ND-GAr HPgTPC with 1~t fiducial mass for 1 year of running in $\nu$-mode, where the horn current direction generates a beam that is primarily neutrinos. The upper portion of the table shows overall numbers of events in broad categories of charged and neutral current interactions, while the lower portion shows event yields for exclusive final states.}
\begin{tabular}{l|c}
\hline
Event class     &     Number of events per ton-year \\
\hline\hline
$\nu_{\mu}$ CC & $1.6 \times 10^{6}$ \\
$\bar{\nu}_{\mu}$ CC & $7.1 \times 10^{4}$ \\
$\nu_{e} + \bar{\nu}_{e}$ CC & $2.9 \times 10^{4}$ \\
NC total & $5.5 \times 10^{5}$ \\ \hline
$\nu_{\mu}$ CC$0\pi$ & $5.9 \times 10^{5}$ \\
$\nu_{\mu}$ CC$1\pi^{\pm}$ & $4.1 \times 10^{5}$ \\
$\nu_{\mu}$ CC$1\pi^{0}$ & $1.6 \times 10^{5}$ \\
$\nu_{\mu}$ CC$2\pi$ & $2.1 \times 10^{5}$ \\
$\nu_{\mu}$ CC$3\pi$ & $9.2 \times 10^{4}$ \\
$\nu_{\mu}$ CC other & $1.8 \times 10^{5}$ \\
\hline \hline
\end{tabular} 
 \label{tab:Stats}
\end{table}

\subsection{Potential to Employ Different Targets}

One unique advantage of ND-GAr, compared to the other DUNE ND components, is its flexibility to use various gas mixtures as interaction targets. A safe and hydrogen-rich gas mixture (such as 92\% Ar and 8\% CH$_{4}$), with the help of the superb tracking, would enable measurements of event-by-event neutrino-hydrogen interactions~\cite{Lu:2015hea,Hamacher-Baumann:2020ogq}, providing direct access to the more fundamental physics parameters, such as the axial form factor and proton radius, for the first time in 40 years, with better detector technology and better understood fluxes.  These measurements would also provide an anchor to enable comparison of Ar/H and C/H cross section ratios. Filling ND-GAr with these other targets could be considered after sufficient data to perform the oscillation analysis constraints are collected with the baseline argon gas mixture.


\subsection{Physics Beyond the Standard Model}

LBNF's high-intensity proton beam will provide a large neutrino flux that will be sampled by ND-GAr.  This will enable DUNE to discover new particles and unveil new interactions and symmetries beyond those predicted in the Standard Model (beyond the standard model, or BSM). In particular, ND-GAr can search for neutrino tridents, heavy neutral leptons (HNL), light dark matter, heavy axions, and anomalous tau neutrinos that come from short-baseline mixing with sterile neutrinos~\cite{DUNE:2020fgq}.

In general, the background contributions in searches for rare events tend to scale directly with the detector mass (which is larger for ND-LAr), while signal events often scale with detector volume (which is similar for ND-LAr and ND-GAr). As a result, the ND-LAr suffers more significantly from backgrounds than ND-GAr. ND-GAr's relatively large volume will be beneficial in the search for the rare decay events. Independent ND-GAr analyses that complement those in ND-LAr will serve to constrain backgrounds and achieve a stronger BSM physics reach with the near detector complex than could be achieved with either detector alone~\cite{koppetalpaper}. 

\subsubsection{Neutrino Tridents}
Neutrino trident production---the production of a pair of oppositely-charged leptons through the scattering of a neutrino on a heavy nucleus---is a powerful probe of new physics in the leptonic sector. The Standard Model expectation is that the DUNE near detector will collect around a dozen of these rare events per ton of argon per year, enough to measure with precision the cross sections of such processes~\cite{Altmannshofer:2019zhy}. The main challenge is to distinguish the trident events from the copious backgrounds, mainly consisting of charged-current single-pion production events, as muon and pion tracks can be easily confused. ND-GAr will will tackle this search by improving muon-pion separation through $dE/dx$ measurements in the HPgTPC and the calorimeter system, and ND-GAr's magntic field will significantly improve signal background separation by tagging the opposite charges of the two leptons in the final state. 

Measurement of neutrino trident interactions may reveal new evidence for or constrain BSM physics models. For instance, a class of models that modify the trident cross section are those that contain an additional neutral gauge boson, $Z^{\prime}_0$, that couples to neutrinos and charged leptons. This $Z^{\prime}_0$ boson can be introduced by gauging an anomaly-free global symmetry of the Standard Model, with a particular interesting case realized by gauging L$_{\mu} -$L$_{\tau}$. Such a $Z^{\prime}_0$ is not very tightly constrained and could address the observed discrepancy between the Standard Model prediction and measurements of the anomalous magnetic moment of the muon, (g$-$2)$_{\mu}$. Its presence would be apparent through a large increase (a factor of 2-3) of the measured trident cross section over the SM prediction, whereas a measurement consistent with the SM would provide new  $Z^{\prime}_0$ constraints in the region of parameter space relevant to (g$-$2)$_{\mu}$.

\subsubsection{Heavy Neutral Leptons}
Searches for several possible HNL decay channels, including those with $\nu e^{+} e^{-}$, $\nu e^{+/-} \mu^{-/+}$, $\nu \mu^{+} \mu^{-}$, $\nu \pi^{0}$, $e^{+/-} \pi^{-/+}$, and $\mu^{+/-} \pi^{-/+}$ final states, can be undertaken in the near detector complex~\cite{breitbach2021searching}. Neutrino interactions such as neutral-current neutral pion (NC $\pi^{0}$) interactions can act as backgrounds to some of these HNL decay channels. In ND-GAr, we will be able to reject many of these backgrounds, since the pressurized gaseous argon medium of ND-GAr can track charged particles with kinetic energies as low as 5~MeV, providing hadronic information right at the interaction vertex and enabling rejection of a significant fraction of the NC $\pi^{0}$ background events. The ND-GAr calorimeter surrounding the argon medium can be used to efficiently reconstruct photons that do not convert in the gas volume, aiding in background rejection for HNL searches. Another example of a dominant background (particularly for HNL channels with $\nu \mu^{+} \mu^{-}$, and $ \mu^{+/-} \pi^{-}$ final states) consists of charged-current muon neutrino interactions with charged pions in the final state, where the muons are misidentified as pions. The pressurized gaseous argon medium in ND-GAr will enable us to more extensively sample ionization per unit track length providing improved muon-pion separation through $dE/dx$ measurements, as can be seen in Fig.~\ref{fig:pep4-dedx}. For even better muon-pion separation, we can further use the calorimeter system in ND-GAr.  In addition to PID in the pressurized gas argon component of ND-GAr, the calorimeter and the muon tagger system will help to further reduce muon-pion confusion, especially at high energies~\cite{DUNE:2021tad}.

\begin{figure}[ht]
\centering
\includegraphics[width=0.6\textwidth]{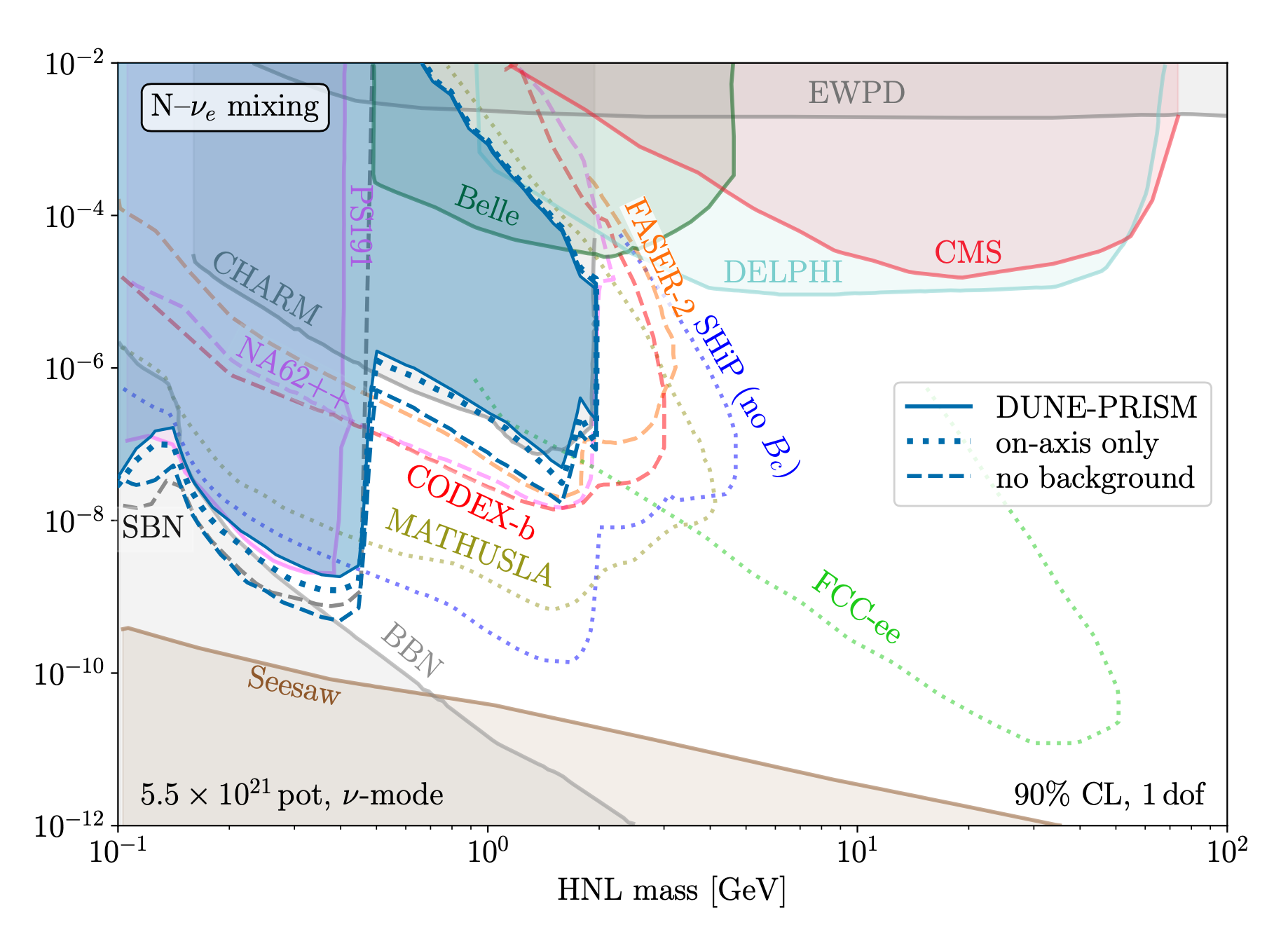}
\caption[DUNE HNL sensitivity]{The background-rejection capability of the ND-GAr detector results in improved coverage of parameter space in HNL searches, as shown by the comparison of the ``no background'' sensitivity with the ``on-axis only'' and ``DUNE-PRISM'' sensitivities~\cite{koppetalpaper}.}
\label{fig:koppetalfig}
\end{figure}

As an example, the DUNE sensitivity to the search for HNL is shown as a function of HNL mass in Figure~\ref{fig:koppetalfig}, which is taken from Ref.~\cite{koppetalpaper}. This sensitivity was obtained by taking into account the individual capabilities of ND-LAr and ND-GAr. In particular, the analysis benefits from the better signal-to-background ratio achieved in ND-GAr, which is due to the fact that HNL signal scales with volume while backgrounds scale with mass. Allowing the detectors to move to positions off the beam axis with the DUNE-PRISM technique only results in a small change to the sensitivity (comparing the ``on-axis'' and ``DUNE-PRISM'' lines in the plot) because effective background suppression can be achieved with ND-GAr even in the on-axis position. Furthermore, since going off-axis also does not harm the sensitivity, despite the lower fluxes in off-axis positions, the search for HNLs can be carried out for any positioning of the detectors~\cite{koppetalpaper}. The ``DUNE-PRISM'' line is the expected achievable sensitivity with ND-LAr and ND-GAr, assuming running time equally split between the on-axis position and six different off-axis locations, while the ``no background'' line represents the sensitivity for a hypothetical background-free analysis, showing that the expected achievable sensitivity at the DUNE near detector complex is not far from what could be achieved with an ``ideal'' experiment.





\subsubsection{Light Dark Matter}
ND-GAr can also search for light dark matter that is produced in the neutrino beam target and subsequently interacts or decays inside the detector. In particular, a light dark matter signal with a single-electron final state (DM e $\rightarrow $ DM e) can be captured with the near detector complex. The neutrino-electron scattering ($\nu$ e $\rightarrow $ and $\nu$ e) and electron neutrino charged current quasi-elastic scattering interactions ($\nu_{e}$ n $\rightarrow $ e p) can look very similar to these signals and can act as backgrounds. According to a study done in Ref.~\cite{lightdm}, the $\nu_{e}$ n $\rightarrow $ e p is easily reducible using energy and angular information of the single-electron final state in ND-LAr. In ND-GAr, we can take advantage of the low detection threshold and use the vertex information to reduce this background even further. From the same reference~\cite{lightdm}, however, the electron scattering interactions are irreducible in ND-LAr, while in ND-GAr, the number of electron scattering events is very small to start with (135 per kton per year, as shown in Table~\ref{tab:Stats}), enabling a nearly background-free light dark matter selection in ND-GAr.  

\subsubsection{Heavy Axions}
There are also opportunities to search for heavy axions in the near detector complex~\cite{PhysRevD.103.095002}. Two main signatures of an axion-like particle have $\gamma \gamma$ and hadrons in final states. The NC $\pi^{0}$ events act as backgrounds to these axion-like particle signals, but as mentioned above, we can easily reject most of these backgrounds by focusing on the rich hadronic information at the vertex that only ND-GAr can provide.

\subsubsection{Tau Neutrinos}
ND-GAr also presents the opportunity to carry out a physics program geared towards probing anomalous $\nu_{\tau}$ interactions. 
The $\tau$ lepton is not directly observable in the DUNE detectors due to its short $2.9\times^{-13}$\,s lifetime, and it is only produced for interactions where the incoming $\nu_\tau$ has an energy of $\sim3.5$\,GeV due to the relatively large $\tau$ mass of 1776.82\,MeV. However, the final state of $\tau$ decays, $\sim65\%$ into hadrons, $\sim18\%$ into $\nu_\tau+e^-+\bar{\nu}_e$, and $\sim17\%$ into $\nu_\tau+\mu^-+\bar{\nu}_\mu$, is readily identifiable in the DUNE ND, given the excellent spatial and energy resolution of ND-GAr and the other ND instruments. While within a three-flavor scenario, the DUNE ND baseline is far too short for $\nu_\mu\rightarrow\nu_\tau$ oscillations to occur, there exists the exciting possibility that $\nu_\tau$ originating in sterile-neutrino driven fast oscillations could be detected with ND-GAr. In particular, ND-GAr would provide excellent sensitivity to the $\tau\rightarrow\mu$ detection channel, with high energy muons in the final state, inaccessible with ND-LAr due to its lack of magnetic field and limited muon energy containment. Furthermore, the TMS detector option is limited to measuring muons up to $\sim$6 GeV/$c$ before they range out, while ND-GAr's use of curvature in the magnetic field to reconstruct muon momentum extends that range to 15 GeV/$c$ and beyond. This extension will become even more valuable when operating LBNF in the high-energy tune, aimed at enriching the available sample of $\nu_{\tau}$ at the far detector, while enhancing sensitivity to anomalous $\nu_{\tau}$ appearance at the near detector~\cite{DeGouvea:2019kea}. Preliminary studies using LBNF's nominal flux and including ND-GAr, estimate that DUNE's sensitivities to anomalous $\nu_\tau$ appearance may extend beyond those of previous searches, such as those carried out by NOMAD and CHORUS.

\section{Detector Performance Requirements} \label{sec:detperf}


The overall DUNE requirements for the near detectors include transferring measurements to far detector, constraining the cross section model, measuring the neutrino flux, and obtaining measurements with different fluxes.   The ND must have reconstruction capabilities that are comparable to or exceed those of the FD in order to effectively transfer measurements.  The ND must measure outgoing recoil particles in neutrino-Ar interactions to ensure that sensitive phase space is properly modeled. It must also measure the wrong sign contamination and measure the intrinsic beam $\nu_{e}$ component.  

ND-GAr is able to meet these requirements with its argon target, magnetization,  uniform angular acceptance, excellent particle identification, low energy tracking thresholds, and wider range for muon momenta measurements. It will also have the ability to probe different off-axis fluxes with spectra spanning region of interest as part of the PRISM concept.  Fulfilling the DUNE ND requirements leads to a set of derived detector capabilities for the ND-GAr.  These are extensively detailed in Ref.~\cite{DUNE:2021tad}, and the key performance capabilities are summarized below.

{\bf Derived ND-GAr detector capabilities}
\begin{itemize}
  
\item The DUNE near detector must classify interactions and measure outgoing particles in a LArTPC with performance comparable to or exceeding that of the FD. It must
measure particles in neutrino-argon interactions with uniform acceptance, lower thresholds than LArTPC, and minimal secondary interaction effects.  To achieve these goals, the ND-GAr must be able to \textbf{constrain the muon energy scale with an uncertainty of 1\% or better} to achieve the oscillation sensitivity described in Vol.~II of the DUNE FD TDR~\cite{DUNE:2020ypp}. The strongest constraint comes from the calibrated magnetic field of the HPgTPC coupled with {\it in-situ} measurements of strange decays. 

  \item {The DUNE ND must be able to measure muons with \textbf{a momentum resolution of less than 4\%}, with non-Gaussian tails corresponding to an RMS $<$ 10\%.  Simulations indicate that a high-pressure gaseous argon TPC should achieve or exceed this resolution.~\cite{DUNE:2021tad} }
  
  \item {The near detector must be able to detect, identify, and measure the momentum of protons emitted from neutrino-argon interactions. To fulfill this, the ND must have a tracking threshold low enough to \textbf{measure the energy spectrum of protons emitted due to final state interactions (FSI) in CC interactions}. Theoretical studies, such as those reported in Refs.~ \cite{Nieves:2005rq,Lalakulich:2012gm,Mosel:2013fxa}, suggest that FSI cause a dramatic increase in final state nucleons with kinetic energies in the range of a few tens of MeV. ND-GAr is suitable for measuring such low-energy protons. The threshold in ND-GAr is an interplay between the argon gas density, readout pixel size, and ionization electron dispersion. Performance studies indicate that a tracking threshold of 5 MeV (or a momentum of 97 MeV/c) is achievable and satisfies this requirement. }
  
  \item {The ND must be able \textbf{to characterize the charged pion energy spectrum in $\nu_{\mu}$ and $\bar{\nu}_{\mu}$ CC interactions from a few GeV down to the low energy region} where FSI are expected to have their largest effect.
      \begin{itemize}
      \item {Theoretical studies, such as those reported in Ref.~\cite{Mosel:2014lja}, predict that FSI are expected to cause a large increase in the number of pions with kinetic energies between 20-150 MeV and a decrease in the range 150-400 MeV. A kinetic energy of 20 MeV corresponds to a momentum of 77 MeV/c. ND-GAr must be able to measure 70 MeV/c charged pions with an efficiency of at least 50\% so as to keep the overall efficiency for measuring events with three pions at the 70 MeV/c threshold above 10\%. 
      }

      \item  {ND-GAr must also have the ability to \textbf{measure the pion multiplicity} and charge in final states with up to 3 pions, so as to inform the pion mass correction in the ND and FD LArTPC. This capability is most important for pions with an energy above a few 100 MeV since those pions predominantly shower in LAr. 
      }
      \end{itemize}
    }
    
    \item {ND-GAr must be able \textbf{to detect and measure $\pi^{0}$'s}, using their decay photons, over the same momentum range as for charged pions. 
    }
    
     \item {ND-GAr must be able \textbf{to identify electrons, muons, charged pions, charged kaons, and protons}. ND-GAr addresses this requirement using a combination of $dE/dx$ in the HPgTPC, $E/p$ using the energy measured in the calorimeter, and the momentum measured by magnetic spectroscopy in the HPgTPC, and by penetration through the calorimeter and muon system. 
    }

\end{itemize}

ND-GAr, because of the calorimeter surrounding its HPgTPC, is also able to characterize the energy carried by neutrons with kinetic energies in the range 50-700~MeV. This capability, while not a requirement, is desirable for constraining uncertainties related to the multiplicity and energies of nucleons ejected from the nucleus during neutrino interactions. 


\section{ND-GAr Detector Overview} \label{sec:det}

\begin{figure}[ht]
\centering
\includegraphics[width=0.5\textwidth]{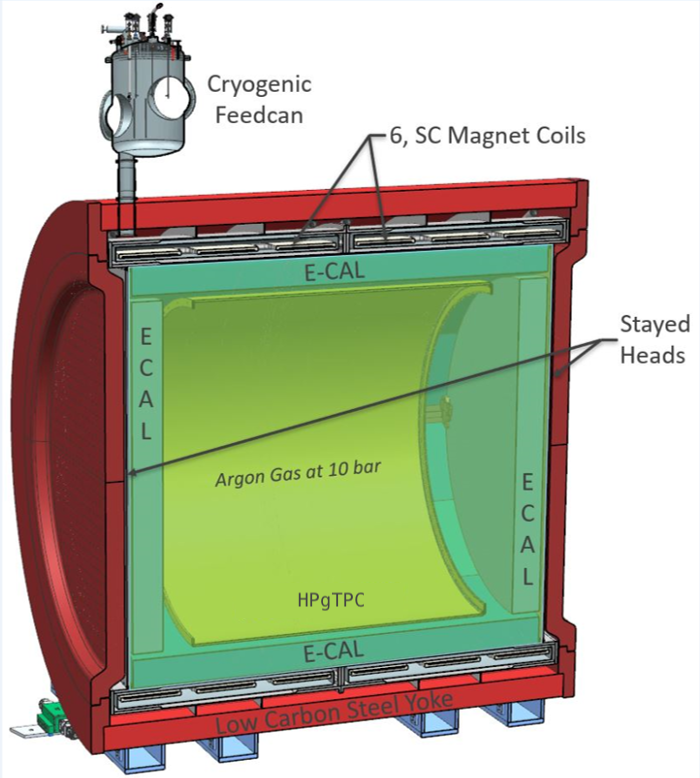}
\caption[ND-GAr Schematic]{Cutaway view of the full ND-GAr detector system showing the HPgTPC, the calorimeter, the magnet, and the iron yoke.  The detectors for the muon-tagging system are not shown.}
\label{fig:ndgarlayout}
\end{figure}

The ND-GAr concept is based on a central high-pressure gaseous argon TPC; the HPgTPC is surrounded by a calorimeter, with both situated in a 0.5~T magnetic field generated by superconducting coils. A muon system is integrated with the magnet return yoke.   A cutaway view of the full ND-GAr system is shown in Fig.~\ref{fig:ndgarlayout}.

The baseline concept of the ND-GAr HPgTPC is based closely on the design of the ALICE TPC~\cite{ALICE:2000jwd}. However, an opportunity exists to reconsider this design
This paper focuses on the baseline design in order to present physics motivations and performance. Future R\&D lines aimed at optimizing this design will be discussed later in the document.  

\subsection{High-Pressure Gaseous Argon Time Projection Chamber and Calorimeter}

At the heart of the ND-GAr detector, the HPgTPC contains an argon-based gas mixture held at a pressure of approximately 10~atm to increase the rate of neutrino interactions.  The drift region of the HPgTPC has a diameter of roughly 5~m and a length of roughly 5~m, corresponding to a fiducial mass of nearly 1~ton of argon. With this fiducial mass, a one-year exposure (defined as $1.1 \times 10^{21}$ protons on target) would result in approximately 1 million neutrino interactions on argon in the on-axis position.  

The calorimeter is inspired by the CALICE analog hadron calorimeter (AHCAL)~\cite{CALICE:2010fpb}. The scintillating layers will consist of a mix of tiles and strips. 
More details of the HPgTPC and calorimeter for the ND-GAr are described in Ref.~\cite{DUNE:2021tad}.

\subsection{Magnet and Pressure Vessel}

The design of the magnet system has evolved over the past few years and no longer uses a separate pressure vessel as was described in Ref.~\cite{DUNE:2021tad}.   The superconducting magnet uses a semi-continuous thin solenoid approach with 6 separate windings.  The design is based on the decades-long evolution of internally wound, aluminum stabilized superconducting magnets.  The required field in the warm bore is relatively low, 0.5T, therefore a single-layer coil is sufficient to provide the needed current density even with a diameter of $\simeq 7$~m.  The design parameters are conservative when compared to previously built magnets.  The iron yoke must be asymmetric to reduce the material budget between ND-LAr and ND-GAr. The magnet system design, a solenoid with a partial return yoke (SPY), makes its design rather unique.

\begin{figure}
    \centering
    \includegraphics[width=0.55\textwidth]{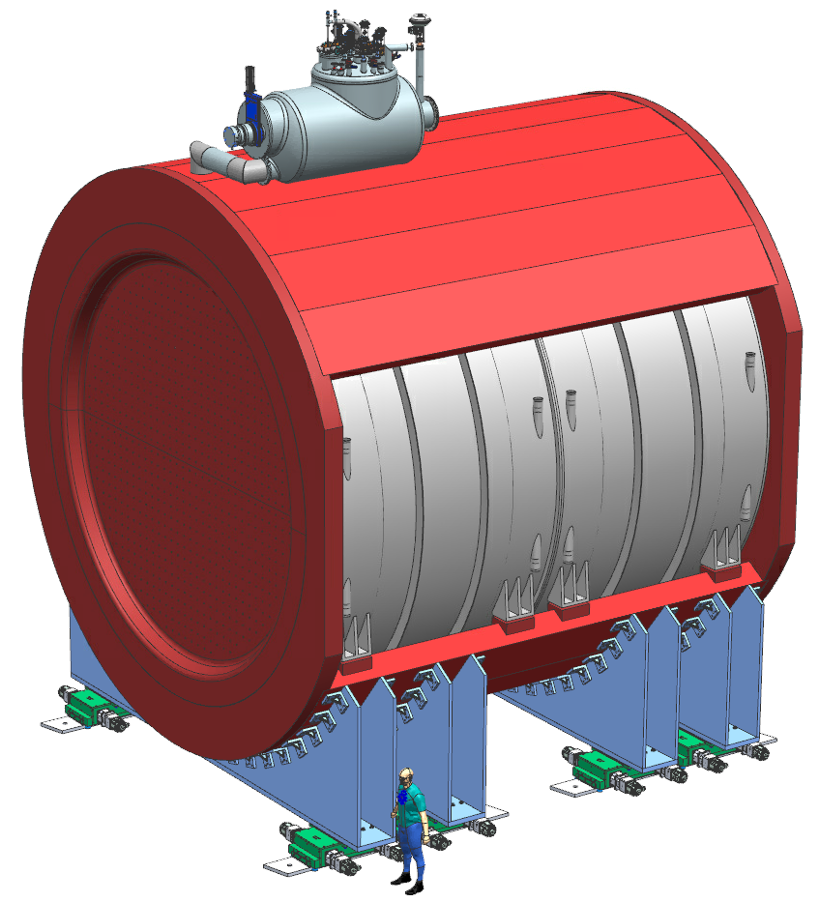}
    \includegraphics[width=0.35\textwidth]{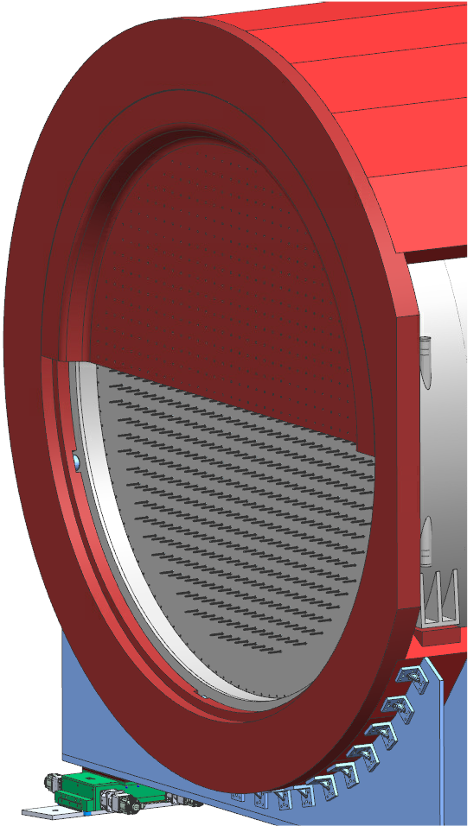}
    \caption{The left image shows the full system, with the magnet return yoke shown in red. The right image shows a cutaway view of one of the end plates, where some of the ``stays'' that support the load of the flat heads are visible.}
    \label{fig:pv}
\end{figure}

A key feature of SPY is that the pressure containment for the 10~bar of argon gas for the HPgTPC is now provided by the solenoid's vacuum vessel and ``stayed'' flat heads are supported by the magnet system yoke.  Figure~\ref{fig:pv} shows the exterior of the complete detector system.  The iron return yoke for the magnet wraps around the sides, top, and bottom of the magnet and has a window on its upstream beam face to minimize the amount of material between the ND-LAr and ND-GAr tracking regions, as shown in Figure~\ref{fig:pv}. The end flanges of the pressure vessel are ``stayed'' by the iron yoke end plates.  

Many of the features of the solenoid are based on the design of the solenoid for the MPD for the Nuclotron-based Ion Collider FAcility (NICA) at JINR~\cite{Abraamyan:2011zz}.

%
\subsection{Muon Tagger}
ND-GAr will need an active layer, the Muon Tagger, outside the iron yoke that will detect particles that penetrate the iron.  These data will be used in conjunction with the calorimeter and HPgTPC to provide $\mu/\pi$ separation. The current design consists of a single active layer outside the iron.  The iron itself is not segmented in depth to allow for multiple measurements. Current technologies under consideration include scintillator extrusions, Resistive Plate Chambers (RPCs), or MicroMegas.

\section{Future R\&D Needs} \label{sec:randd}

While much of the DUNE ND-GAr baseline design is based on the ALICE TPC and CALICE calorimeter designs, there are several important differences and requirements.  The major areas of R\&D needed for ND-GAr are outlined below.

\begin{enumerate}

\item \textbf{Gas mixture studies} 

The ALICE TPC baseline gas mixture is Ne/CO$_2$/N$_2$ at near atmospheric pressure. Neon offers a higher response to ionization than argon (higher avalanche gain) while the combination of a high ion mobility (stemming from the fact that neon is a light gas) and small ageing effects (through the use of CO$_2$ as the quenching gas) are crucial choices at the reaction rates typical of LHC operation. When aiming at high gain operation of argon gas in a low-rate environment, however, an Ar/hydrocarbon-based mixture is \emph{a priori} preferred, as was done in PEP-4 (Ar/CH$_4$, 80/20). For ND-GAr, establishing the optimum quencher concentration will require a balance between the present safety limits of the mixture (e.g. 8\% for CH$_4$), minimizing the number of non-Ar interactions, maintaining good drift-diffusion characteristics, and allowing sufficient charge gain~\cite{Hamacher-Baumann:2020ogq}. A promising R\&D line is the addition of wavelength-shifting species in the gas phase, which would allow time-tagging of that fraction of events for which the charged-particle content is not energetic enough to be time-tagged in the calorimeter.

\item \textbf{Light Collection} 

Primary light production in pure argon in the vacuum ultraviolet (wavelengths $<$200 nm) is well understood~\cite{Gonzalez-Diaz:2017gxo}. In pure argon at a pressure of 10~atm, we estimate that a minimum ionizing particle will produce approximately 400 photons/cm~\cite{Chandrasekharan:2005edc}, but in typical gas TPC operation, a quenching gas or gases are added that either quench or absorb most of the VUV emission.   Recent studies have indicated that with the addition of Xe~\cite{Gonzalez-Diaz:lidine2021}
or CF$_4$~\cite{Gonzalez-Diaz:APS2021} (among possibly other wavelength-shifting species), it is possible to quench the hard VUV component of argon (thus improving avalanche gain) while producing light in the visible or near-IR bands. In Ref.~\cite{Gonzalez-Diaz:APS2021} in particular, the possibility of producing up to 700 photons/MeV in Ar/CF$_4$ mixtures at 1\% molar fraction is shown. With suitable instrumentation (e.g., a SiPM plane at the cathode and a teflon reflector in the field cage) about 350~photoelectrons/5MeV track are anticipated. Utilizing this light would be a novel development for a gaseous argon TPC.
However, R$\&$D is needed in order to understand the impact of impurities on the yields, maximum avalanche gain achievable in different charge amplification structures, mitigation of secondary scintillation backgrounds, and finally the design of a photon detection system. In close coordination with groups investigating the gas mixture, field cage, HV, and DAQ, a conceptual design will be developed for the collection and readout of light in the gas volume, once a suitable gas mixture is identified.

\item \textbf{Readout Chambers}  

Figure~\ref{fig:HPgTPC_ROC} indicates the locations of Inner Readout Chambers (IROCs) and Outer Readout Chambers (OROCs) in ALICE. These chambers have been removed from ALICE as part of the upgrade for LHC Run 3, and can be used for DUNE.   At a minimum, even in the baseline scenario, new  readout chambers will need to be designed to cover the central area of the endcaps, which was not part of the TPC in ALICE.  

However, it is possible to consider a totally new design for the full circular readout plane. This could employ a different technology, such as GEM or MicroMegas charge readout.  Such charge amplification structures allow staging, thereby being more flexible in principle at achieving high and stable gains. In the case of using a scintillating mixture like Ar/CF$_4$ for reconstructing the primary scintillation (t$_0$) signal, readouts based on Gas Electron Multipliers (GEMs) can be optimized for screening of the secondary scintillation signal into the photosensor plane.

Alternatively, and given the present technological landscape, it seems timely to consider a fully-optical readout, as a scintillating gas mixture is already under investigation for the $t_0$ light collection system. For this readout concept, a camera is mounted on the outside of the active gas volume and images, through a transparent anode, the scintillation light produced during the gas amplification of tracks. This produces 2D images of the interactions inside of the TPC. Optical readout could offer an unprecedented pixel size in the readout plane equivalent to $\leq1\times$~mm$^2$ over an area as large as $1$~m$^2$, where all pixels are read out by one integrated device, the camera. It would also offer the attractive feature of passing signals from millions of readout channels through the pressure vessel wall at just the cost of a window. The third coordinate perpendicular to the anode plane can be reconstructed by using fast cameras or tailored ones such as the TimePix3 camera, which have already demonstrated effective pixel sizes down to $3\times3$ mm$^2$~\cite{Fisher_Levine_2016, Roberts_2019}. 
 
ALICE also had a central cathode with two-drift regions for the electrons, with the readout chambers placed on both ends of the detector. The main advantages of this design are the simplicity of HV insulation, about a factor two reduction in purity requirements and collection times, a reduction of space-charge effects, and a factor $\sqrt{2}$ reduction in ionization spread. Alternatively, DUNE could consider a single-drift region, with the readout on one side, which would reduce the number of electronics channels needed, and would provide a space for a light-collection system near the cathode. The optimization of the chamber readout must be developed in close connection with performance figures such as tracking and energy threshold, $dE/dx$, energy and momentum resolution, and physics impact of a t$_0$ signal.

\begin{figure}[htb]
\centerline{ \includegraphics[width=0.5\textwidth]{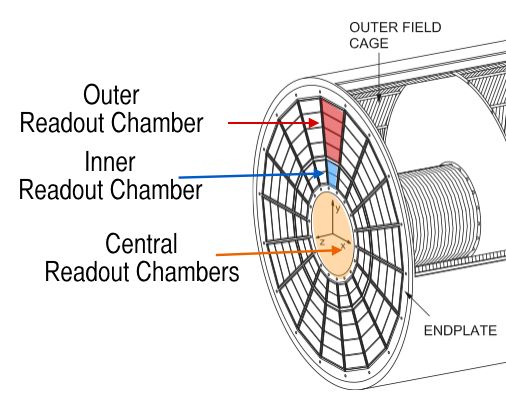}}
\caption{Diagram of a possible arrangement of readout chambers of the ND-GAr HPgTPC, based on a drawing of Ref.~\cite{Alme:2010ke}. The Central Readout Chambers would need to be designed and added to the HPgTPC.}
\label{fig:HPgTPC_ROC}
\end{figure}

\item \textbf{Electronics and DAQ development} 


The electronic readout of the detector is under development with an aim to maximize synergies with existing electronics. The system's design is driven by the high magnetic field and pressure that the electronics must be situated in. The primary limitation of the pressure is that data and power cables must pass through the pressure vessel, so some in-vessel aggregation is necessary to limit the total number of penetrations. The planned system is comprised of front end cards (FECs) hosting digitizer ASICs; these FECs are read out by a smaller number of in-vessel aggregation cards (AGGs), which then send data to out-of-vessel timing, interface and power cards (TIPs). We aim to digitize the signal to the level of tens of MHz, and this is driven by z-axis resolution.
For the FEC, two approaches are under investigation. The first uses similar electronics to the liquid argon near detector, based on the LArPix ASIC~\cite{dwyerLArPixDemonstrationLowpower2018a} chip. This chip would achieve very low cost per channel, but some modifications are needed to adapt it for use in the HPgTPC, since the HPgTPC signal is faster and inverted compared to the liquid argon near detector (multi-wire-proportional-chamber-based gaseous argon readout chambers mainly detect induced charge from ions moving in relation to the pickup electrodes, rather than electrons moving to collection plane wires), and the gain in the gas results in a widened dynamic range. The other approach is based on the SAMPA ASIC~\cite{Barboza:2016ala} in use by ALICE. This system has a higher cost per channel but does not require modification. The AGGs and TIPs are being designed to both use the same PCB design to minimize costs, and will both be based around the same Xilinx Kintex FPGA. Both power and data will be passed down the same cables, with the current plan to use standard ethernet cables and RJ45 feedthroughs.

Readout electronics will also need to be developed for a light collection system.

\item \textbf{Field Cage and High Voltage}

A new field cage and mechanical endcap structures will need to be constructed for the ND-GAr HPgTPC. Since the ND-GAr detector will be movable, a robust mechanical design for the field cage will be needed. A buffer region with an insulating gas is planned for the region in between the field cage and pressure vessel.  As the HPgTPC will be operated at higher pressure and cathode potential, the insulating gas and vessel-field cage spacing requires optimization~\cite{Norman:2021fdy}.  Differential-pressure regulation between the different gas volumes will be needed. 

\item \textbf{Calibration} 

To precisely monitor any variations of the drift velocity and inhomogeneities in the drift field,  a laser calibration system  that can illuminate the entire drift volume is desirable.  Its design will need to be developed in close collaboration with the high voltage field cage design, and R\&D will need to be performed to ensure that an adequate signal amplitude can be obtained in the high-pressure gas mix. 



\item \textbf{Calorimeter}  

The calorimeter must fulfill several roles, ranging from photon energy and direction measurements to muon/pion separation and neutron detection, which impose different constraints on the capabilities of the system. While the general technology of plastic scintillator tiles and strips read out with silicon photomultipliers is already well-established, most notably in the framework of CALICE\cite{CALICE:2010fpb}, several areas of R\&D specific to ND-GAr remain. On the system level, this includes the design and optimization of the absorber geometry, readout granularity and of the overall mechanical structure. Specific front-end electronics that are suitable for the high envisioned channel count, provide a time resolution on the few 100 picosecond level, and possibly the capability for pulse-shape discrimination while meeting the low-power requirements, will need to be developed. In the area of scintillators, strip concepts that provide a sufficient time resolution for the measurement of neutron energies via time of flight, are needed. Specific scintillator materials that enhance the neutron detection capability, and their matching to silicon-based photon sensors, will need to be studied.

\item \textbf{Muon system}  

The ND-GAr muon system is in a very preliminary stage of design, and it crucially depends on the particulars of both the calorimeter and magnet systems.  The current design consists of a single layer outside the iron.  The iron itself is not segmented in depth to allow for multiple measurements.  Current technologies under consideration include scintillator extrusions, resistive plate chambers, or MicroMegas.

\end{enumerate}

\section{Summary} \label{sec:conclusion}

The complete ND-GAr detector is required in order for DUNE to reach its full physics potential.  A magnetized tracker will allow for excellent momentum resolution and sign-selection for the long-baseline oscillation physics.  Its wide acceptance and low tracking threshold for neutrino interactions on argon will provide crucial contributions to the oscillation physics.
ND-GAr will also provide a rich physics program of cross-section measurements and will extend the capabilities of the DUNE ND for physics beyond the standard model.  While the current design benefits from work done on the ALICE TPC and by the CALICE calorimeter collaboration, it is desirable both to adapt it to the DUNE ND environment, and to explore new technological developments that offer a window of opportunity. A focused R\&D program will need to be implemented, aimed at completion of the detector baseline as well as exploration of these timely new technologies and extended physics reach.

\section*{Acknowledgements} \label{sec:acknowledgements}

This document was prepared by the DUNE collaboration using the
resources of the Fermi National Accelerator Laboratory 
(Fermilab), a U.S. Department of Energy, Office of Science, 
HEP User Facility. Fermilab is managed by Fermi Research Alliance, 
LLC (FRA), acting under Contract No. DE-AC02-07CH11359.
%
%
This work was supported by
CNPq,
FAPERJ,
FAPEG and 
FAPESP,                         Brazil;
CFI, 
IPP and 
NSERC,                          Canada;
CERN;
M\v{S}MT,                       Czech Republic;
ERDF, 
H2020-EU and 
MSCA,                           European Union;
CNRS/IN2P3 and
CEA,                            France;
INFN,                           Italy;
FCT,                            Portugal;
NRF,                            South Korea;
CAM, 
Fundaci\'{o}n ``La Caixa'',
Junta de Andaluc\'ia-FEDER,
MICINN, and
Xunta de Galicia,               Spain;
SERI and 
SNSF,                           Switzerland;
T\"UB\.ITAK,                    Turkey;
The Royal Society and 
UKRI/STFC,                      United Kingdom;
DOE and 
NSF,                            United States of America.

\bibliographystyle{JHEP}
\bibliography{references.bib}


\newpage

\end{document}